\documentclass[PRL,aps,nofootinbib,longbibliography,preprint,twocolumn,two-sided,
superscriptaddress,10pt,showpacs]{revtex4-1}
\usepackage{graphicx}% Include figure files
\usepackage{color}
\usepackage{xfrac}
\usepackage{amsmath,amssymb}
\usepackage{fixmath}

\usepackage[english]{babel}
\usepackage[autostyle, english = british]{csquotes}
\MakeOuterQuote{"}
\usepackage{textgreek}
\usepackage[export]{adjustbox}
\usepackage{siunitx}
\usepackage{xcolor}
\usepackage{hyperref}
\usepackage{lipsum}
\usepackage{multirow}
\usepackage{notes2bib}
\usepackage[version=4]{mhchem} % Chemical compound names
\usepackage{gensymb} % For \degree (°) symbol
\usepackage{xfrac} % For slanted fractions

\newcommand{\Tc}{\ensuremath{T_\mathrm{c}}}
\newcommand{\etal}{\textit{et al.} }

\newcommand{\nno}{$[110]_\mathrm{HTT}$}
\newcommand{\noo}{$[100]_\mathrm{HTT}$}
\newcommand{\oon}{$[001]_\mathrm{HTT}$}
\newcommand{\ono}{$[010]_\mathrm{HTT}$}
\newcommand{\lco}{\ce{La2CuO4}}
\newcommand{\lmo}{\ce{La2MgO4}}

\DeclareUnicodeCharacter{2212}{\ensuremath{-}}

\ifdefined\qty\else
  \ifdefined\NewCommandCopy
    \NewCommandCopy\qty\SI
  \else
    \NewDocumentCommand\qty{O{}mm}{\SI[#1]{#2}{#3}}
  \fi
\fi
\ifdefined\unit\else
  \ifdefined\NewCommandCopy
    \NewCommandCopy\unit\si
  \else
    \NewDocumentCommand\unit{O{}m}{\si[#1]{#2}}
  \fi
\fi

\begin{document}
% Header of the manuscript
% Title, authors and abstract

% Title
\title{The origin of strain-induced stabilisation of superconductivity in the
lanthanum cuprates}

% List of authors
\author{Christopher Keegan}
\affiliation{Departments of Physics and Materials, and the Thomas Young Centre,
Imperial College London, London SW7 2AZ, United Kingdom}

\author{Mark S. Senn}
\affiliation{Department of Chemistry, University of Warwick, Gibbet Hill,
Coventry, CV4 7AL, United Kingdom}

\author{Nicholas C. Bristowe}
\affiliation{Centre for Materials Physics, Durham University, South Road,
Durham, DH1 3LE, United Kingdom}

\author{Arash A. Mostofi}
\email{a.mostofi@imperial.ac.uk}
\affiliation{Departments of Physics and Materials, and the Thomas Young Centre,
Imperial College London, London SW7 2AZ, United Kingdom}

\date{\today}

\begin{abstract}
Suppression of superconductivity in favour of a striped phase, and its
coincidence with a structural transition from a low-temperature orthorhombic
(LTO) to a low-temperature tetragonal (LTT) phase, is a ubiquitous feature of
hole-doped lanthanum cuprates.
We study the effect of anisotropic strain on
this transition using density-functional theory on both \lco\ and the
recently-synthesised surrogate \lmo\ to decouple electronic and structural
effects.
Strikingly, we find that compressive strain applied diagonally to the
in-plane metal-oxygen bonds dramatically stabilises the LTO phase.
Given the mutual exclusivity of 3D superconductivity and long-range static
stripe order, we thereby suggest a structural mechanism for understanding
experimentally-observed trends in the superconducting \Tc\ under uniaxial
pressure, and suggest principles for tuning it.
\end{abstract}

\maketitle

\section{Introduction}
\begin{figure}
	\centering
  \includegraphics[width=\columnwidth]{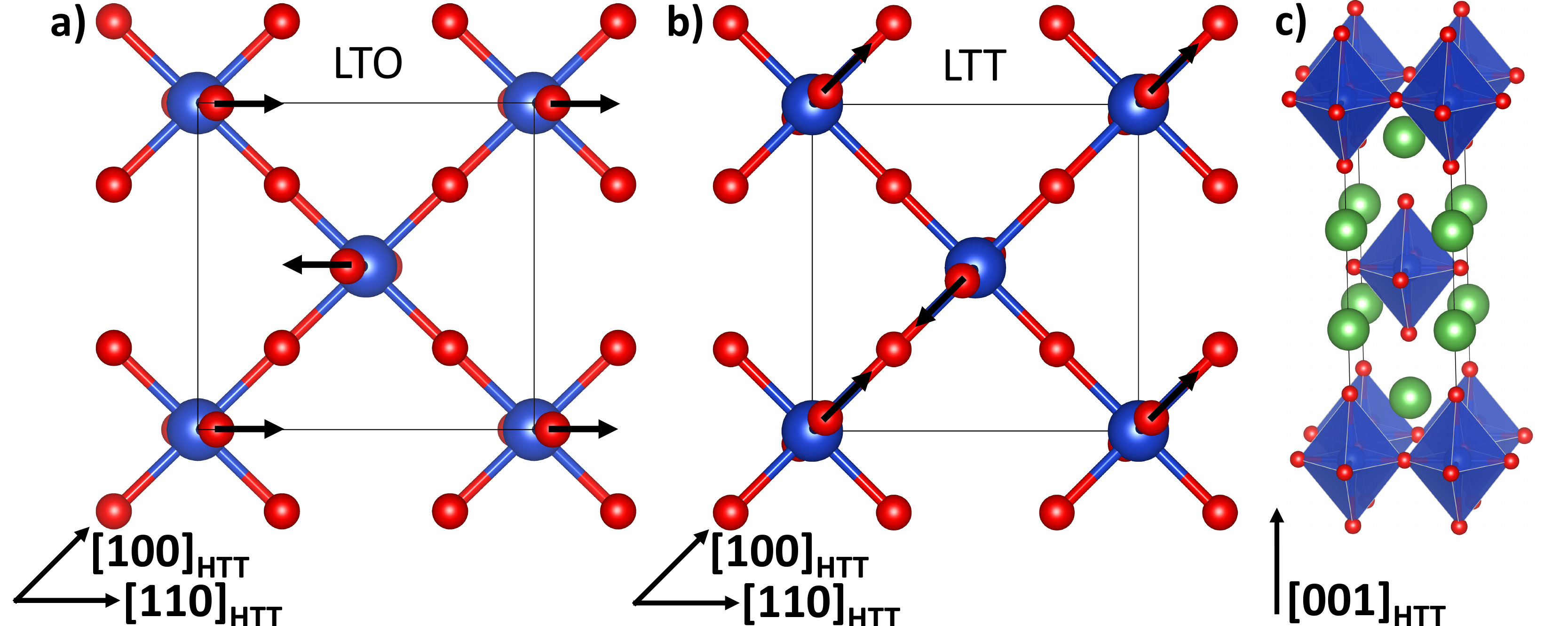}
  \caption{
     {\bf a-b)} Layering-axis view of the \ce{CuO2} planes in the {\bf a)} low-temperature
     orthorhombic (LTO) and {\bf b)} low-temperature tetragonal (LTT) phases
     of \ce{La2CuO4}. The black arrows indicate the
     displacement of the apical oxygen
     associated with the octahedral tilt, which is at \ang{45} to and along
     the \ce{Cu-O} bond in the LTO and LTT phases, respectively. {\bf c)}
     High-temperature tetragonal (HTT) phase with untilted \ce{CuO6} octahedra.
     The \nno, \noo\ and \oon\ directions, along which we apply uniaxial strain, are indicated.
  }
  \label{fig:1}
\end{figure}

Since the discovery of a superconducting transition temperature of
$\Tc=27$~\unit{\kelvin} in \ce{La_{2-x}Ba_{x}CuO4} (LBCO) \cite{Bednorz1986}, 
the lanthanum cuprates have become the paradigmatic class of materials for 
studying \mbox{high-\Tc} cuprates. Lanthanum cuprates undergo
a second-order structural phase transition from a high-temperature tetragonal
(HTT) $I4/mmm$ phase, to a low-temperature orthorhombic (LTO) $Bmab$ phase
\cite{Paul1987} in which the \ce{CuO6} octahedra
tilt along an axis at \ang{45} to the in-plane \ce{Cu-O} bonds, as shown in
Fig.~\ref{fig:1}a.

In the case of \ce{La_{2-x}A_{x}CuO4} (A = Ba, Nd/Sr or Eu/Sr)
and $x \simeq \sfrac{1}{8}$, there is an additional phase transition from
LTO to a low-temperature tetragonal (LTT) $P4_2/ncm$ phase
\cite{Axe1989, Cox1989, Crawford1991}.
The LTT phase features an octahedral tilt along the \ce{Cu-O} bonds (Fig.~\ref{fig:1}b), which
alternates between the \noo\ and \ono\ directions in adjacent layers
\cite{Suzuki1989,Axe1989}.
In LBCO, for example, the LTO-LTT phase transition coincides with a dramatic drop in \Tc\ \cite{Moodenbaugh1988}, which is attributed to the 
stabilisation of charge-stripe order
\cite{Hucker2011, Hucker2012} that competes with 3D superconductivity \cite{Axe1994, Hucker2011}. 
Experiments have shown that both hydrostatic and anisotropic pressure
influence \Tc\ and stripe-order in the lanthanum cuprates
\cite{Murayama1991, Yamada1992, Katano1993, Arumugam2002, Arumugam2002a,
Takeshita2004, Hucker2010, Guguchia2020, Boyle2021, Guguchia2023}. \Tc\ 
increases with the application of uniaxial pressure in the \ce{CuO2} plane
\cite{Arumugam2002a, Takeshita2004}, and the enhancement is more substantial
when pressure is applied at \ang{45} to the \ce{Cu-O} bond axis as 
compared to parallel to it \cite{Takeshita2004}.
This effect has been attributed to the destabilisation of static stripe order
\cite{Guguchia2020, Tranquada2020, Guguchia2023}, a reduction of the \ce{CuO6}
octahedral tilt angle in the LTT phase \cite{Arumugam2002}, and a suppression
of the LTO-LTT phase transition temperature \cite{Hucker2010, Boyle2021,
Guguchia2023}. 
These interpretations are consistent with recent X-ray scattering
experiments that have measured marked reductions in the temperatures
associated with the LTO-LTT phase transition and onset of static stripe order
when uniaxial pressure is applied along \noo\ \cite{Boyle2021}.

From the perspective of theory, the importance of strong electron correlations
in the lanthanum cuprates poses
significant challenges for understanding the physics of these systems and, of
particular relevance to this work, the underlying mechanisms behind the
experimental observations described above.

Recently, to circumnavigate this problem, we synthesised and studied a novel
structural surrogate of the lanthanum cuprates, namely \lmo~\cite{Tidey2022}.  
\lmo\ is an insulator that does not exhibit anti-ferromagnetic order,
Jahn-Teller distortions, or the charge-transfer band gap 
associated with strongly correlated electronic phases in the lanthanum cuprates.
Conveniently, \ce{Mg^{2+}} (0.73~\AA) and \ce{Cu^{2+}} (0.72~\AA) have almost
identical ionic radii \cite{Shannon1976}, and \lmo\ exhibits the same series of
structural phase transitions (HTT-LTO-LTT) as the lanthanum cuprates
\cite{Tidey2022}. This indicates that the LTO-LTT phase transition can
present itself independently of any electronic or magnetic order, and
provides an opportunity to investigate the mechanism underlying this
phase transition in a system in which strong electron correlation is absent.

In this work, we investigate the effect of uniaxial strain
on the structural phase stability of \ce{La2MgO4} and \ce{La2CuO4} using
first-principles calculations based on density-functional theory (DFT).
We show that uniaxial compressive strain in the \ce{MgO2} plane energetically
stabilises the LTO phase relative to the LTT phase. This effect is
enhanced when the strain direction is at \ang{45} to the \ce{Mg-O} bond
directions (\nno) as compared to when it is parallel (\noo).
This contrasts with uniaxial strain along the layering axis (\oon), which has
the opposite effect. We find that \ce{La2CuO4} exhibits qualitatively the same behaviour. 
We compare our calculated stabilisation energies to experimental measurements
of \Tc\ in a lanthanum cuprate under different uniaxial pressure conditions
\cite{Takeshita2004} and show that there is a remarkable correlation. 
We suggest that the mechanism underlying this correlation may be due to phase
coexistence of LTO and LTT in experiments, whereby the phase fraction of LTO increases
with increasing in-plane uniaxial pressure and, hence, enables the development of
3D superconductivity in the sample.

\section{Methods}

\subsection{Total energy calculations}
We calculate the ground-state atomic and electronic structures of \ce{La2BO4}
(B = Mg, Cu) within density-functional theory as implemented in the Quantum
ESPRESSO plane-wave pseudopotential software package \cite{QE-2009, QE-2017}. 
We use the PBEsol \cite{Perdew2008} generalised gradient approximation (GGA) to
describe exchange and correlation, and ultrasoft pseudopotentials from the
GBRV library \cite{Garrity2014}. 
We apply a Hubbard $U$ of 9~\unit{\electronvolt}
to Cu $3d$ orbitals and of 4~\unit{\electronvolt} to La $4f$ orbitals.
We use a 60~Ry energy cutoff for the plane-wave basis expansion of the
Kohn-Sham eigenstates, with an eight-times higher cutoff energy to represent
the electronic charge density and augmentation regions associated with the
ultrasoft pseudopotentials. 
For LTO and LTT, we use  $\sqrt{2}\times\sqrt{2}\times1$ 
real-space supercells of the HTT parent conventional unit cell.
In the case of LTO and LTT, we sample the Brillouin zone with a regular
Monkhorst-Pack mesh of dimensions $5\times5\times2$ for \ce{La2MgO4} and
$8\times8\times4$ for \ce{La2CuO4}. For HTT, we use $8\times8\times4$ and
$10\times10\times4$, respectively.
All atomic positions and lattice vectors are fully relaxed, with total energies
converged to within 10$^{-9}$~Ry, ionic forces to within
10$^{-5}$~Ry\,$a_{0}^{-1}$ and stresses to within 5~MPa. 
We apply uniaxial strain by varying the relevant lattice parameter,
and constraining it while allowing all other
degrees of freedom to relax. We define strain, $\epsilon$, as

\begin{equation}
\epsilon = \frac{\xi - \xi_0}{\xi_0},
\label{eq:strain}
\end{equation}

where $\xi$ is the relevant lattice parameter, and 
$\xi_0 = \frac{1}{2}(a_{0}^{\text{LTO}}+b_{0}^{\text{LTO}})$ or $c_{0}^{\text{LTO}}$
is the average of the in-plane, or the $c$-axis, lattice parameters of the
fully-relaxed LTO phase. Further details of the computational methods can be
found in Sec.~S1 of the Supplementary Information \cite{SI_LBO2022}.

\subsection{Symmetry analysis}
We describe the relationship of the low-symmetry LTO and LTT phases relative to
the parent HTT phase using a two-dimensional order parameter (OP) transforming
as the X$_3^+$(a;b) irreducible representation (irrep). The OP corresponds to the octahedral
\ce{BO6} tilt. The OP magnitude is the norm of the X$_3^+$(a;b)
irrep, and the OP direction is defined as the angle $\phi = \arctan(b/a)$.
OP directions of \ang{0} (X$_3^+$(a;0)) or \ang{90} (X$_3^+$(0;a)) correspond
to the LTO phase, and an OP direction of \ang{45} (X$_3^+$(a;a)) corresponds
to the LTT phase. Intermediate OP directions correspond to a structure
consistent with space-group symmetry of the $Pccn$ subgroup.
We determined space-group symmetry using FINDSYM
\cite{FINDSYM_Stokes2005}, and irreps associated with the OP using ISODISTORT
\cite{ISODISPLACE_Campbell2006} from the ISOTROPY software suite. We
visualised crystal structures in CIF format using the VESTA 3D visualisation
program \cite{VESTA_2011} and used it to create the images in Fig.~\ref{fig:1}.

\section{Results}

\subsection{Ground state structure and energetics of \ce{La2BO4} (B = Cu, Mg)}

For both \lmo\ and \lco, our calculated lattice parameters, unit cell volumes,
\ce{B-O} and \ce{La-O} bond lengths are in good agreement 
with experimental values measured at 10~K
(within 1.5\%, 1.8\% and 2\%, respectively) \cite{Radaelli1994, Tidey2022}.
\lmo\ exhibits a 4.4\% contraction of the layering axis, and a 7.8\%
contraction in the axial octahedral bond length as compared to \lco. 
Further, we compute the magnitude of the Jahn-Teller $Q_3$ modes as 
\begin{equation}
  Q_3 = \frac{2(l -s)}{\sqrt{2}},
  \label{eqn:q3}
\end{equation}
where $l$ and $s$ are the axial and equatorial octahedral bond lengths, respectively.
We find that the $Q_3$ mode in \lco\ (LTO) is 0.71~\AA\ as compared to
0.35~\AA\ in \lmo\ (LTT). These values match corresponding $Q_3$ modes in
experimental $I4/mmm$ structures for \lco\ (ICSD ref. 41643 \cite{Zhang1992}:
$Q_3$ = 0.74~\AA) and \lmo\ (Tidey \etal\ 900~K structure \cite{Tidey2022}:
\mbox{$Q_3$ = 0.43~\AA}).
This demonstrates the ability of our calculations to capture the Jahn-Teller
distortion in \lco.
We find that in both \lmo\ and \lco\ the lowest-energy phase is LTT,
followed by LTO and then HTT.
In \lmo\ the energy difference between the LTO and LTT phases
is 11.7~meV per formula unit (f.u.), whereas in \lco\ it is much smaller at
2.4~meV/(f.u.), as also reported previously \cite{Tidey2022}.
Related DFT calculations on \lco\ using the SCAN meta-GGA
functional for exchange and correlation have similarly found a very small energy 
difference between the LTT and LTO phases of 1.2~meV/(f.u.)~\cite{Furness2018} and less than 0.5~meV/(f.u.)~\cite{Pokharel2022}.
Although the lowest energy structure in these previous calculations is the LTO
phase, the very similar energies of LTO and LTT suggests that these
phases are closely competing.

\subsection{Effect of uniaxial strain on the ground-state structure of
\ce{La2MgO4}}

\begin{figure}
	\centering
  \includegraphics[width=\columnwidth]{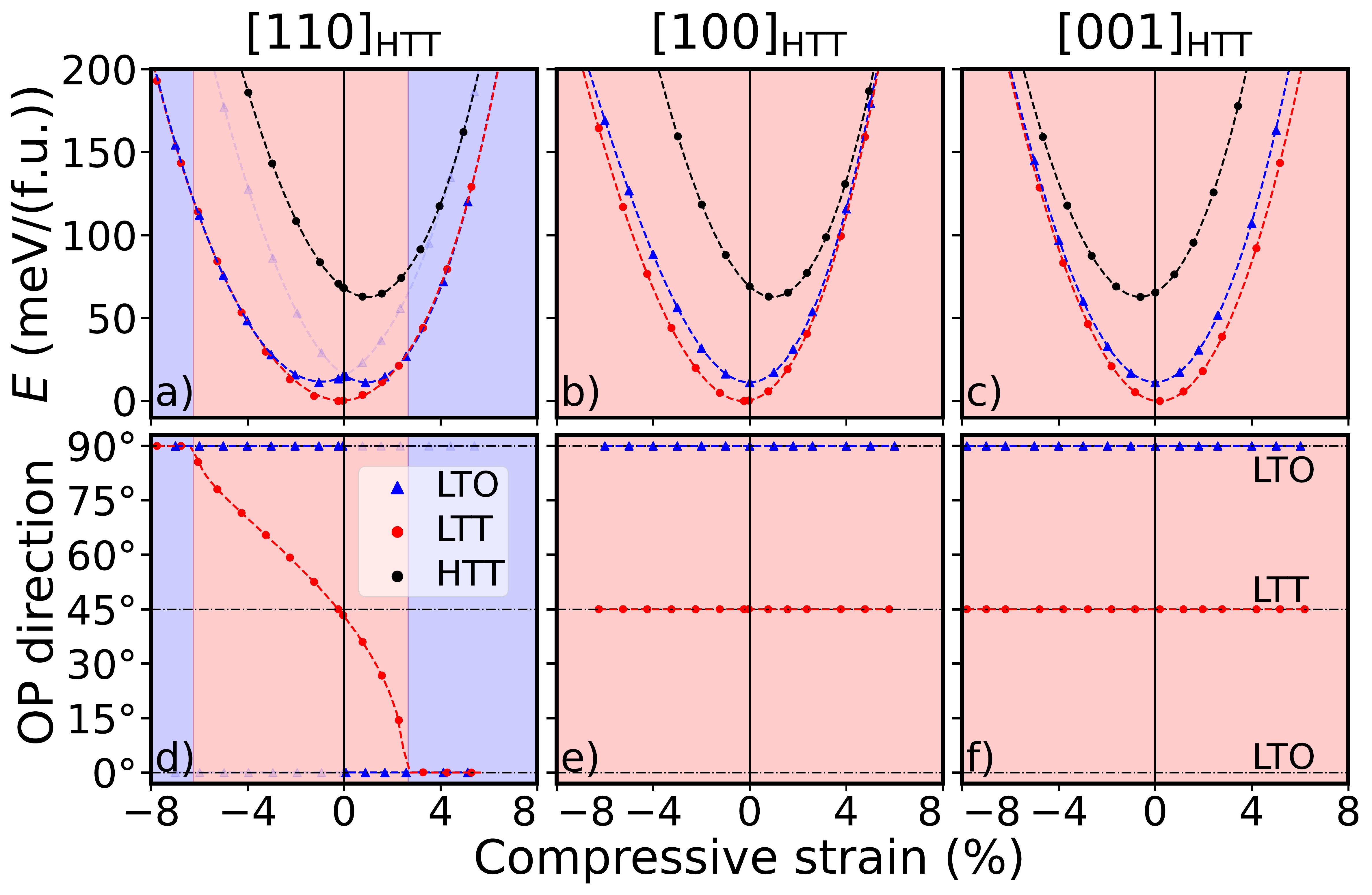}
  \caption{
    a-c) Total energy, relative to the relaxed 
  LTT phase, of the LTO,
  LTT and HTT phases of \lmo\ as a function of
  uniaxial compressive strain along the a) \nno, b) \noo\ and c) \oon\ axes.
  d-f) Order parameter (OP)
  direction as a function of uniaxial compressive strain along the d) \nno, e)
  \noo\ and f) \oon\ axes. OP directions of \ang{0} and \ang{90} correspond to the
  LTO phase, and of \ang{45} to the LTT phase. The shading indicates the lowest
  energy phase at each strain, either LTT (red) or LTO (blue).
  }
  \label{fig:2}
\end{figure}

Figure~\ref{fig:2}a-c) compares the total energies of the HTT (black), LTT
(red) and LTO (blue) phases of \lmo\ as a function of compressive uniaxial
strain along the \nno, \noo\ and \oon\ axes.
In the case of strain along the
\nno\ axis (Figure~\ref{fig:2}a), there are two curves for the LTO phase, 
which correspond to strain applied along the largest and
smallest in-plane axis; the opaque blue curve 
highlights the lowest energy LTO phase at each given strain.
Further details of the
implementation of the strained total-energy calculations is given in
Sec.~S1 of the Supplementary Information \cite{SI_LBO2022}.
It can be seen that compressive strain in the out-of-plane direction (i.e.,
along \oon) further destabilises the LTO phase relative to LTT. Conversely,
compressive uniaxial in-plane strain (i.e., along \nno\ and \noo) has the
opposite effect and stabilises the LTO phase. 

Figures~\ref{fig:2}d-f) show the order parameter (OP) direction as a function
of compressive strain for the LTT and LTO phases.
Angles of \ang{0} and \ang{90} correspond to the LTO phase, and \ang{45}
corresponds to the LTT phase. 
It can be seen that only strain along \nno\ causes the
OP direction of the LTT phase to change, which happens progressively until it
converges to \ang{0} (\ang{90}) under compressive (tensile) strain. In this
strain regime, the space-group symmetry of the LTT phase changes to $Pccn$,
which allows the OP to rotate continuously towards the LTO phase.
Strain along either \noo\ or \oon\ does not change the OP direction of either
the LTO or LTT phase. 
The background shading in Figure~\ref{fig:2} indicates the lowest energy phase
for each given value of strain. 

\begin{figure}
	\centering
  \includegraphics[width=\columnwidth]{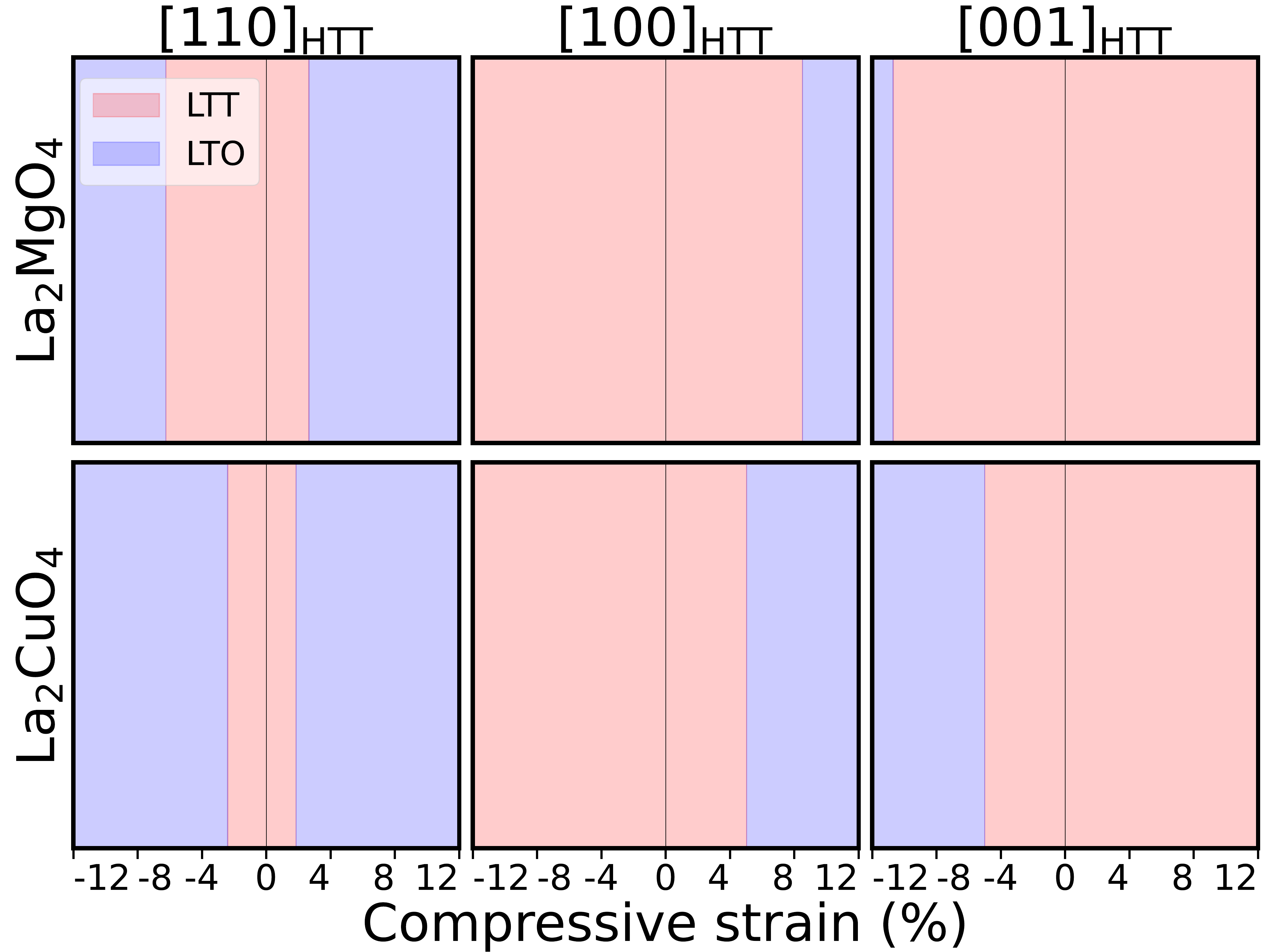}
  \caption{Regions of stability of the LTT (red) and LTO (blue) phases of \ce{La2BO4}
  (B = Cu, Mg) as a function of uniaxial compressive strain along the
  \nno, \noo\ and \oon\ axes. 
  The data underlying the \lco\ phase diagram is shown in Figure S1 in the
  Supplementary Information \cite{SI_LBO2022}.
  }
  \label{fig:3}
\end{figure}

In Figure~\ref{fig:3}, we compare the regions of stability of the HTT, LTT and LTO
phases of \lmo\ and \lco\ under the three uniaxial strain regimes.
We find that the behaviour of \lco\ is qualitatively very similar to that of
\lmo, highlighting the similar response to strain in both systems, and thus the
ability of the \lmo\ to serve as a useful structural surrogate to the
lanthanum cuprates. 
The magnitude of the strain at which the LTO phase is stabilised in
\lco\ is smaller than in \lmo, and this can be understood from the
much smaller energy difference between the LTO and LTT phases in
unstrained \lco\ as compared to unstrained \lmo.

\begin{figure}
	\centering
  \includegraphics[width=\columnwidth]{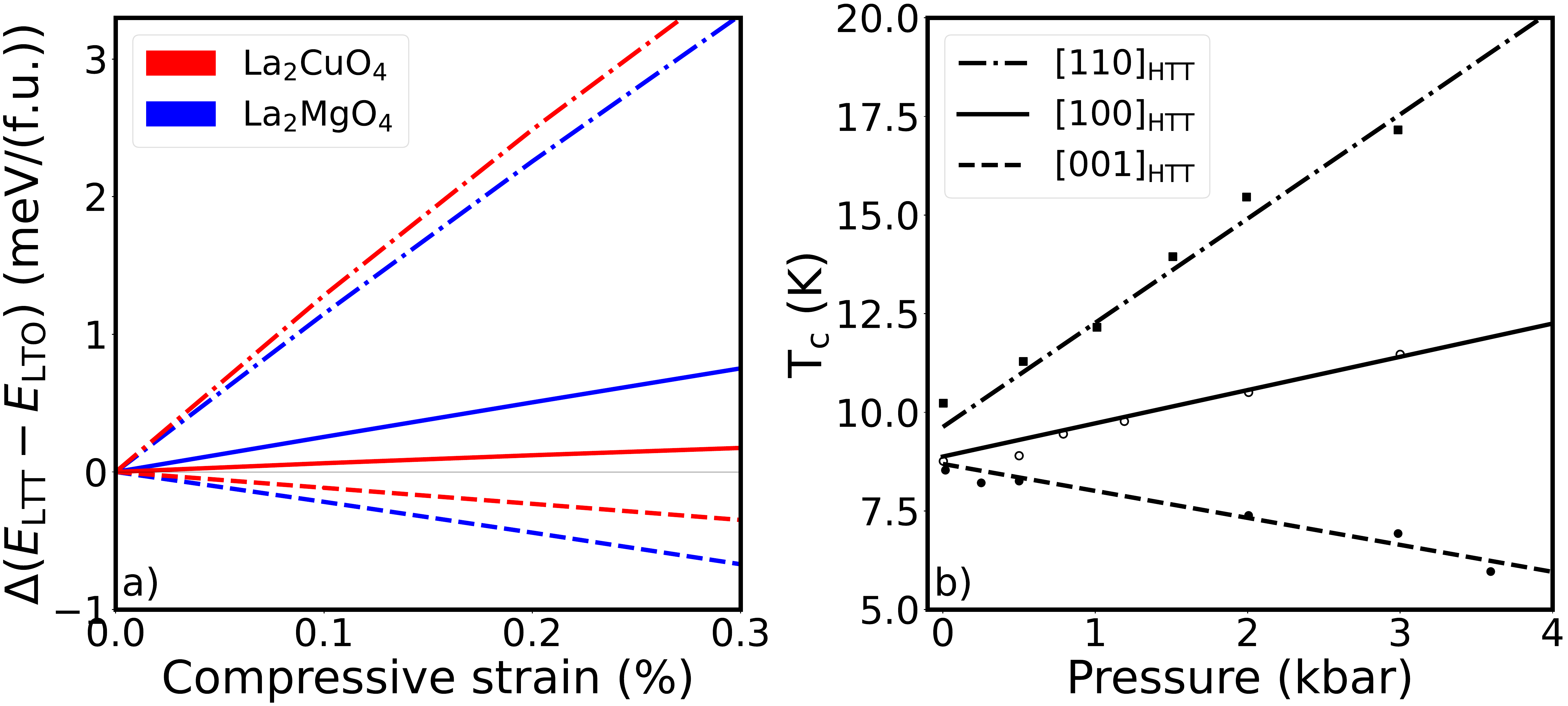}
  \caption{a) The change in the energy difference between the LTT and LTO phases
  as a function of compressive strain in \ce{La2BO4} (B = Cu, Mg) for uniaxial
  strain applied along the \nno, \noo\ and \oon\ axes.
  b) Variation in superconducting \Tc\ as a
  function of uniaxial pressure applied along the \nno,
  \noo\ and \oon\ axes in \ce{La_{1.64}Eu_{0.2}Sr_{0.16}CuO_4}. 
  Data replotted from \cite{Takeshita2004}.
  }
  \label{fig:4}
\end{figure}

Fig.~\ref{fig:4}a shows the change in relative stability of the LTT and LTO
phases for \lmo\ (blue lines) and \lco\ (red lines) as a function of compressive
uniaxial strain along the \nno\ (dot-dashed lines), \noo\ (solid lines) and
\oon\ (dashed lines) axes. 
It can be seen that out-of-plane uniaxial compressive strain destabilises the
LTO phase relative to the LTT phase. Conversely, however, in-plane uniaxial
compressive strain stabilises the LTO phase and the effect is more significant
for strain along the \nno\ direction than for the \noo\ direction. 
We compare these trends to measurements of the superconducting \Tc\ in
\ce{La_{1.64}Eu_{0.2}Sr_{0.16}CuO4} as a function of compressive pressure along
the same three axes, conducted by Takeshita~\etal\cite{Takeshita2004} and
reproduced in Fig.~\ref{fig:4}b, noting that 4~kbar corresponds approximately
to 0.24\% compressive strain in these systems.
There is an evident correlation between the change in relative stability of the
LTT and LTO phases as a function of compressive strain and the evolution of \Tc.
A natural question arises: what is the origin of this correspondence?

Takeshita~\etal\cite{Takeshita2004}, assuming their sample to be
entirely composed of the LTT phase, interpreted their result as the manifestation
of strong coupling between uniaxial strain and stripe order. They reasoned that
stripes, which form along the \ce{Cu-O} bond direction, with a \ang{90} rotation
of the stripe
direction in adjacent \ce{CuO2} planes, are more effectively destabilised by
uniaxial pressure at \ang{45} to the stripe direction (i.e., along
\nno) than along it (i.e., \noo), since the latter is more commensurate with
local stripe symmetry. 
The destabilisation of stripe order 
renders it less effective at competing with the superconducting state
for the ground state, hence the observed increase in \Tc, particularly
when pressure is applied along \nno.
Very recently, however, we have shown~\cite{Tidey2022a} that
in both LBCO ($x = 0.125$) and LESCO ($x = 0.125$, $y = 0.12$), significant 
phase coexistence persists between LTT and LTO (or $Pccn$) domains down to
temperatures as low as 10~K. It is likely, therefore, that the sample studied 
by Takeshita~\etal\cite{Takeshita2004} also exhibited significant phase
coexistence.
Indeed, it is not unreasonable to expect a significant LTO phase fraction
for the sample to exhibit the reported 3D superconductivity, and
H\"ucker~\etal\cite{Hucker2010} also suggested that the sample studied by
Takeshita~\etal may have undergone a phase transition from LTT to LTO. 
In light of these observations, we suggest an alternative explanation for the
enhancement of \Tc\ under uniaxial pressure that is based on phase coexistence,
and is consistent with our calculated energetics of the LTT and LTO phases as a
function of uniaxial strain. 
We assume a sample that has coexisting LTT and LTO grains, and that 3D
superconductivity arises as a result of phase coherence developing between
patches of LTO phase~\cite{Tranquada2020}.
It has been found that 2D superconductivity arises at the onset temperature
of static spin-stripe order \cite{Tranquada2008}, but that the phase coherence
of the interlayer Josephson coupling is frustrated due to the alternating stripe
direction in the LTT phase \cite{Tajima2001, Tranquada2020}. This prevents 3D
superconductivity from evolving in the latter phase.
When in-plane uniaxial compressive strain is applied, the LTO phase is
stabilised with respect to LTT and, hence, the equilibrium phase fraction of
LTO should be expected to increase. 
As the phase fraction of LTO phase increases, the average separation of the
LTO regions decreases and, hence, superconducting phase coherence develops
more easily, which manifests itself as an observed increase in \Tc.
This scenario is consistent with recent results that demonstrate a significant
decrease in the LTO-LTT phase transition temperature for a 
\ce{La_{1.48}Nd_{0.4}Sr_{0.12}CuO4} sample \cite{Boyle2021}, and an increase
in 3D superconducting \Tc \cite{Guguchia2020} when uniaxial stress is applied. 
Tranquada \cite{Tranquada2020} considered a similar scenario to explain the 
occurrence of superconductivity in LBCO ($x = \sfrac{1}{8}$) at low
temperatures. Tranquada considered that coherence could develop between dilute
patches of uniform $d$-wave order, giving rise to 3D superconductivity.
In Tranquada’s scenario, the dilute patches are ascribed to
doping inhomogeneity \cite{Tranquada2020}. In our
scenario, we view these dilute patches as being related to the LTO and LTT
phase coexistence and domain structure. It is of course entirely possible that
LTO/LTT domains nucleate preferentially around areas of
charge inhomogeneity, meaning these scenarios are not mutually exclusive.

In conclusion, we have investigated the effect of uniaxial strain on the 
structural phase stability of \lmo\ and \lco\ using first-principles
density-functional theory calculations. Our results show clearly that
compressive uniaxial strain applied in the \ce{BO2} (B = Mg, Cu) plane
stabilises the LTO phase relative to LTT, and that the effect is most dramatic
when the strain is applied at \ang{45} to the \ce{B-O} bonds. Strikingly, we
find a strong positive correlation between our calculated stabilisation of the
LTO phase and
previous experimental measurements of superconducting \Tc\ in the lanthanum
cuprates under uniaxial pressure. We have proposed a
mechanism to explain this correlation that is based upon phase coherence
developing between patches of superconducting LTO phase that are in
coexistence with the LTT phase. In-plane compressive uniaxial strain stabilises
the LTO phase, increases its volume phase fraction, and thereby permits 3D
superconductivity to persist to higher temperatures.
Our results point to a general strategy for tuning 
superconductivity in this class of materials.

\subsection{Note on strain-pressure conversion}

We justify the comparison of uniaxial pressure and compressive strain in
Fig.~\ref{fig:4} by noting that the in-plane Young's modulus of
\lco\ at 44~K is $\sim$167~GPa~\cite{Migliori1990}.
Assuming linear elasticity, a uniaxial stress of 4~kbar (0.4~GPa) would then
induce a compressive strain of approximately 0.24\%.

\section*{Data Availability}
Input and output files for all calculations used in this work, or 
support our findings have been made available in a figshare repository 
and can be accessed using the digital object identifier: 
\url{doi.org/10.6084/m9.figshare.21647381}.

\nocite{apsrev41Control}
\bibliographystyle{apsrev4-1} 
\newpage
\bibliography{references.bib}	

\begin{thebibliography}{39}%
\makeatletter
\providecommand \@ifxundefined [1]{%
 \@ifx{#1\undefined}
}%
\providecommand \@ifnum [1]{%
 \ifnum #1\expandafter \@firstoftwo
 \else \expandafter \@secondoftwo
 \fi
}%
\providecommand \@ifx [1]{%
 \ifx #1\expandafter \@firstoftwo
 \else \expandafter \@secondoftwo
 \fi
}%
\providecommand \natexlab [1]{#1}%
\providecommand \enquote  [1]{``#1''}%
\providecommand \bibnamefont  [1]{#1}%
\providecommand \bibfnamefont [1]{#1}%
\providecommand \citenamefont [1]{#1}%
\providecommand \href@noop [0]{\@secondoftwo}%
\providecommand \href [0]{\begingroup \@sanitize@url \@href}%
\providecommand \@href[1]{\@@startlink{#1}\@@href}%
\providecommand \@@href[1]{\endgroup#1\@@endlink}%
\providecommand \@sanitize@url [0]{\catcode `\\12\catcode `\$12\catcode
  `\&12\catcode `\#12\catcode `\^12\catcode `\_12\catcode `\%12\relax}%
\providecommand \@@startlink[1]{}%
\providecommand \@@endlink[0]{}%
\providecommand \url  [0]{\begingroup\@sanitize@url \@url }%
\providecommand \@url [1]{\endgroup\@href {#1}{\urlprefix }}%
\providecommand \urlprefix  [0]{URL }%
\providecommand \Eprint [0]{\href }%
\providecommand \doibase [0]{http://dx.doi.org/}%
\providecommand \selectlanguage [0]{\@gobble}%
\providecommand \bibinfo  [0]{\@secondoftwo}%
\providecommand \bibfield  [0]{\@secondoftwo}%
\providecommand \translation [1]{[#1]}%
\providecommand \BibitemOpen [0]{}%
\providecommand \bibitemStop [0]{}%
\providecommand \bibitemNoStop [0]{.\EOS\space}%
\providecommand \EOS [0]{\spacefactor3000\relax}%
\providecommand \BibitemShut  [1]{\csname bibitem#1\endcsname}%
\let\auto@bib@innerbib\@empty
%</preamble>
\bibitem [{\citenamefont {Bednorz}\ and\ \citenamefont
  {M\"uller}(1986)}]{Bednorz1986}%
  \BibitemOpen
  \bibfield  {author} {\bibinfo {author} {\bibfnamefont {J.~G.}\ \bibnamefont
  {Bednorz}}\ and\ \bibinfo {author} {\bibfnamefont {K.~A.}\ \bibnamefont
  {M\"uller}},\ }\bibfield  {title} {\enquote {\bibinfo {title} {Possible
  high-{T}$_\mathrm{}$ superconductivity in the {B}a-{L}a-{C}u-{O} system},}\
  }\href {\doibase 10.1007/bf01303701} {\bibfield  {journal} {\bibinfo
  {journal} {Zeitschrift f\"ur Physik B -- Condensed Matter}\ }\textbf
  {\bibinfo {volume} {64}},\ \bibinfo {pages} {189--193} (\bibinfo {year}
  {1986})}\BibitemShut {NoStop}%
\bibitem [{\citenamefont {Paul}\ \emph {et~al.}(1987)\citenamefont {Paul},
  \citenamefont {Balakrishnan}, \citenamefont {Bernhoeft}, \citenamefont
  {David},\ and\ \citenamefont {Harrison}}]{Paul1987}%
  \BibitemOpen
  \bibfield  {author} {\bibinfo {author} {\bibfnamefont {D.~M.}\ \bibnamefont
  {Paul}}, \bibinfo {author} {\bibfnamefont {G.}~\bibnamefont {Balakrishnan}},
  \bibinfo {author} {\bibfnamefont {N.~R.}\ \bibnamefont {Bernhoeft}}, \bibinfo
  {author} {\bibfnamefont {W.~I.~F.}\ \bibnamefont {David}}, \ and\ \bibinfo
  {author} {\bibfnamefont {W.~T.~A.}\ \bibnamefont {Harrison}},\ }\bibfield
  {title} {\enquote {\bibinfo {title} {Anomalous {S}tructural {B}ehavior of the
  {S}uperconducting {C}ompound {L}a$_{1.85}${B}a$_{0.15}${C}u{O}$_4$},}\
  }\href@noop {} {\bibfield  {journal} {\bibinfo  {journal} {Physical Review
  Letters}\ }\textbf {\bibinfo {volume} {58}},\ \bibinfo {pages} {1976--1978}
  (\bibinfo {year} {1987})}\BibitemShut {NoStop}%
\bibitem [{\citenamefont {Axe}\ \emph {et~al.}(1989)\citenamefont {Axe},
  \citenamefont {Moudden}, \citenamefont {Hohlwein}, \citenamefont {Cox},
  \citenamefont {Mohanty}, \citenamefont {Moodenbaugh},\ and\ \citenamefont
  {Xu}}]{Axe1989}%
  \BibitemOpen
  \bibfield  {author} {\bibinfo {author} {\bibfnamefont {J.~D.}\ \bibnamefont
  {Axe}}, \bibinfo {author} {\bibfnamefont {A.~H.}\ \bibnamefont {Moudden}},
  \bibinfo {author} {\bibfnamefont {D.}~\bibnamefont {Hohlwein}}, \bibinfo
  {author} {\bibfnamefont {D.~E.}\ \bibnamefont {Cox}}, \bibinfo {author}
  {\bibfnamefont {K.~M.}\ \bibnamefont {Mohanty}}, \bibinfo {author}
  {\bibfnamefont {A.~R.}\ \bibnamefont {Moodenbaugh}}, \ and\ \bibinfo {author}
  {\bibfnamefont {Youwen}\ \bibnamefont {Xu}},\ }\bibfield  {title} {\enquote
  {\bibinfo {title} {Structural phase transformations and superconductivity in
  {L}a$_{2-x}${B}a$_x${C}u{O}$_4$},}\ }\href {\doibase
  10.1103/PhysRevLett.62.2751} {\bibfield  {journal} {\bibinfo  {journal}
  {Physical Review Letters}\ }\textbf {\bibinfo {volume} {62}},\ \bibinfo
  {pages} {2751--2754} (\bibinfo {year} {1989})}\BibitemShut {NoStop}%
\bibitem [{\citenamefont {Cox}\ \emph {et~al.}(1989)\citenamefont {Cox},
  \citenamefont {Zolliker}, \citenamefont {Axe}, \citenamefont {Moudden},
  \citenamefont {Moodenbaugh},\ and\ \citenamefont {Xu}}]{Cox1989}%
  \BibitemOpen
  \bibfield  {author} {\bibinfo {author} {\bibfnamefont {D.~E.}\ \bibnamefont
  {Cox}}, \bibinfo {author} {\bibfnamefont {P.}~\bibnamefont {Zolliker}},
  \bibinfo {author} {\bibfnamefont {J.~D.}\ \bibnamefont {Axe}}, \bibinfo
  {author} {\bibfnamefont {A.~H.}\ \bibnamefont {Moudden}}, \bibinfo {author}
  {\bibfnamefont {A.~R.}\ \bibnamefont {Moodenbaugh}}, \ and\ \bibinfo {author}
  {\bibfnamefont {Y.}~\bibnamefont {Xu}},\ }\bibfield  {title} {\enquote
  {\bibinfo {title} {Structural {S}tudies of {L}a$_{2-x}${B}a$_x${C}u{O}$_4$
  {B}etween 11–293 {K}},}\ }\href {\doibase 10.1557/PROC-156-141} {\bibfield
  {journal} {\bibinfo  {journal} {MRS Proceedings}\ }\textbf {\bibinfo {volume}
  {156}},\ \bibinfo {pages} {141} (\bibinfo {year} {1989})}\BibitemShut
  {NoStop}%
\bibitem [{\citenamefont {Crawford}\ \emph {et~al.}(1991)\citenamefont
  {Crawford}, \citenamefont {Harlow}, \citenamefont {McCarron}, \citenamefont
  {Farneth}, \citenamefont {Axe}, \citenamefont {Chou},\ and\ \citenamefont
  {Huang}}]{Crawford1991}%
  \BibitemOpen
  \bibfield  {author} {\bibinfo {author} {\bibfnamefont {M.~K.}\ \bibnamefont
  {Crawford}}, \bibinfo {author} {\bibfnamefont {R.~L.}\ \bibnamefont
  {Harlow}}, \bibinfo {author} {\bibfnamefont {E.~M.}\ \bibnamefont
  {McCarron}}, \bibinfo {author} {\bibfnamefont {W.~E.}\ \bibnamefont
  {Farneth}}, \bibinfo {author} {\bibfnamefont {J.~D.}\ \bibnamefont {Axe}},
  \bibinfo {author} {\bibfnamefont {H.}~\bibnamefont {Chou}}, \ and\ \bibinfo
  {author} {\bibfnamefont {Q.}~\bibnamefont {Huang}},\ }\bibfield  {title}
  {\enquote {\bibinfo {title} {Lattice instabilities and the effect of
  copper-oxygen-sheet distortions on superconductivity in doped
  {L}a$_2${C}u{O}$_4$},}\ }\href {\doibase 10.1103/PhysRevB.44.7749} {\bibfield
   {journal} {\bibinfo  {journal} {Physical Review B}\ }\textbf {\bibinfo
  {volume} {44}},\ \bibinfo {pages} {7749--7752} (\bibinfo {year}
  {1991})}\BibitemShut {NoStop}%
\bibitem [{\citenamefont {Suzuki}\ and\ \citenamefont
  {Fujita}(1989)}]{Suzuki1989}%
  \BibitemOpen
  \bibfield  {author} {\bibinfo {author} {\bibfnamefont {Takashi}\ \bibnamefont
  {Suzuki}}\ and\ \bibinfo {author} {\bibfnamefont {Toshizo}\ \bibnamefont
  {Fujita}},\ }\bibfield  {title} {\enquote {\bibinfo {title} {Anomalous
  {C}hange in {C}rystalline {S}tructure of
  ({L}a$_{1-x}${B}a$_x$)$_2${C}u{O}$_{4-\delta}$},}\ }\href@noop {} {\bibfield
  {journal} {\bibinfo  {journal} {Journal of the Physical Society of Japan}\
  }\textbf {\bibinfo {volume} {58}},\ \bibinfo {pages} {1883--1886} (\bibinfo
  {year} {1989})}\BibitemShut {NoStop}%
\bibitem [{\citenamefont {Moodenbaugh}\ \emph {et~al.}(1988)\citenamefont
  {Moodenbaugh}, \citenamefont {Xu}, \citenamefont {Suenaga}, \citenamefont
  {Folkerts},\ and\ \citenamefont {Shelton}}]{Moodenbaugh1988}%
  \BibitemOpen
  \bibfield  {author} {\bibinfo {author} {\bibfnamefont {A.~R.}\ \bibnamefont
  {Moodenbaugh}}, \bibinfo {author} {\bibfnamefont {Youwen}\ \bibnamefont
  {Xu}}, \bibinfo {author} {\bibfnamefont {M.}~\bibnamefont {Suenaga}},
  \bibinfo {author} {\bibfnamefont {T.~J.}\ \bibnamefont {Folkerts}}, \ and\
  \bibinfo {author} {\bibfnamefont {R.~N.}\ \bibnamefont {Shelton}},\
  }\bibfield  {title} {\enquote {\bibinfo {title} {Superconducting properties
  of {L}a$_{2-x}${B}a$_x${C}u{O}$_4$},}\ }\href {\doibase
  10.1103/PhysRevB.38.4596} {\bibfield  {journal} {\bibinfo  {journal}
  {Physical Review B}\ }\textbf {\bibinfo {volume} {38}},\ \bibinfo {pages}
  {4596--4600} (\bibinfo {year} {1988})}\BibitemShut {NoStop}%
\bibitem [{\citenamefont {Hücker}\ \emph {et~al.}(2011)\citenamefont
  {Hücker}, \citenamefont {Zimmermann}, \citenamefont {Gu}, \citenamefont
  {Xu}, \citenamefont {Wen}, \citenamefont {Xu}, \citenamefont {Kang},
  \citenamefont {Zheludev},\ and\ \citenamefont {Tranquada}}]{Hucker2011}%
  \BibitemOpen
  \bibfield  {author} {\bibinfo {author} {\bibfnamefont {M}~\bibnamefont
  {Hücker}}, \bibinfo {author} {\bibfnamefont {M~V}\ \bibnamefont
  {Zimmermann}}, \bibinfo {author} {\bibfnamefont {G~D}\ \bibnamefont {Gu}},
  \bibinfo {author} {\bibfnamefont {Z~J}\ \bibnamefont {Xu}}, \bibinfo {author}
  {\bibfnamefont {J~S}\ \bibnamefont {Wen}}, \bibinfo {author} {\bibfnamefont
  {Guangyong}\ \bibnamefont {Xu}}, \bibinfo {author} {\bibfnamefont {H~J}\
  \bibnamefont {Kang}}, \bibinfo {author} {\bibfnamefont {A}~\bibnamefont
  {Zheludev}}, \ and\ \bibinfo {author} {\bibfnamefont {J~M}\ \bibnamefont
  {Tranquada}},\ }\bibfield  {title} {\enquote {\bibinfo {title} {Stripe order
  in superconducting {L}a$_{2−x}${B}a$_x${C}u{O}$_4$ ($0.095 \leq x \leq
  0.155$)},}\ }\href {\doibase 10.1103/PhysRevB.83.104506} {\bibfield
  {journal} {\bibinfo  {journal} {Physical Review B}\ }\textbf {\bibinfo
  {volume} {83}},\ \bibinfo {pages} {104506} (\bibinfo {year}
  {2011})}\BibitemShut {NoStop}%
\bibitem [{\citenamefont {Hücker}(2012)}]{Hucker2012}%
  \BibitemOpen
  \bibfield  {author} {\bibinfo {author} {\bibfnamefont {M.}~\bibnamefont
  {Hücker}},\ }\bibfield  {title} {\enquote {\bibinfo {title} {Structural
  aspects of materials with static stripe order},}\ }\href {\doibase
  10.1016/J.PHYSC.2012.04.035} {\bibfield  {journal} {\bibinfo  {journal}
  {Physica C: Superconductivity}\ }\textbf {\bibinfo {volume} {481}},\ \bibinfo
  {pages} {3--14} (\bibinfo {year} {2012})}\BibitemShut {NoStop}%
\bibitem [{\citenamefont {Axe}\ and\ \citenamefont {Crawford}(1994)}]{Axe1994}%
  \BibitemOpen
  \bibfield  {author} {\bibinfo {author} {\bibfnamefont {J.~D.}\ \bibnamefont
  {Axe}}\ and\ \bibinfo {author} {\bibfnamefont {M.~K.}\ \bibnamefont
  {Crawford}},\ }\bibfield  {title} {\enquote {\bibinfo {title} {Structural
  instabilities in lanthanum cuprate superconductors},}\ }\href {\doibase
  10.1007/BF00754942} {\bibfield  {journal} {\bibinfo  {journal} {Journal of
  Low Temperature Physics}\ }\textbf {\bibinfo {volume} {95}},\ \bibinfo
  {pages} {271--284} (\bibinfo {year} {1994})}\BibitemShut {NoStop}%
\bibitem [{\citenamefont {Murayama}\ \emph {et~al.}(1991)\citenamefont
  {Murayama}, \citenamefont {Tamegai}, \citenamefont {Iye}, \citenamefont
  {Môri}, \citenamefont {Oguro}, \citenamefont {Yomo}, \citenamefont {Takagi},
  \citenamefont {Uchida},\ and\ \citenamefont {Tokura}}]{Murayama1991}%
  \BibitemOpen
  \bibfield  {author} {\bibinfo {author} {\bibfnamefont {Chizuko}\ \bibnamefont
  {Murayama}}, \bibinfo {author} {\bibfnamefont {Tsuyoshi}\ \bibnamefont
  {Tamegai}}, \bibinfo {author} {\bibfnamefont {Yasuhiro}\ \bibnamefont {Iye}},
  \bibinfo {author} {\bibfnamefont {Nobuo}\ \bibnamefont {Môri}}, \bibinfo
  {author} {\bibfnamefont {Isamu}\ \bibnamefont {Oguro}}, \bibinfo {author}
  {\bibfnamefont {Shusuke}\ \bibnamefont {Yomo}}, \bibinfo {author}
  {\bibfnamefont {Hideroni}\ \bibnamefont {Takagi}}, \bibinfo {author}
  {\bibfnamefont {Shinichi}\ \bibnamefont {Uchida}}, \ and\ \bibinfo {author}
  {\bibfnamefont {Yoshinori}\ \bibnamefont {Tokura}},\ }\bibfield  {title}
  {\enquote {\bibinfo {title} {Pressure effect on the transport properties in
  the low temperature phase of {L}a$_{1.875}${B}a$_{0.125}${C}u{O}$_4$},}\
  }\href {\doibase 10.1016/0921-4526(91)90365-L} {\bibfield  {journal}
  {\bibinfo  {journal} {Physica B: Physics of Condensed Matter}\ }\textbf
  {\bibinfo {volume} {169}},\ \bibinfo {pages} {639--640} (\bibinfo {year}
  {1991})}\BibitemShut {NoStop}%
\bibitem [{\citenamefont {Yamada}\ and\ \citenamefont
  {Ido}(1992)}]{Yamada1992}%
  \BibitemOpen
  \bibfield  {author} {\bibinfo {author} {\bibfnamefont {N.}~\bibnamefont
  {Yamada}}\ and\ \bibinfo {author} {\bibfnamefont {M.}~\bibnamefont {Ido}},\
  }\bibfield  {title} {\enquote {\bibinfo {title} {Pressure effects on
  superconductivity and structural phase transitions in
  {L}a$_{2-x}${M}$_x${C}u{O}$_4$ ({M} = {B}a, {S}r)},}\ }\href {\doibase
  10.1016/0921-4534(92)90029-C} {\bibfield  {journal} {\bibinfo  {journal}
  {Physica C: Superconductivity and its applications}\ }\textbf {\bibinfo
  {volume} {203}},\ \bibinfo {pages} {240--246} (\bibinfo {year}
  {1992})}\BibitemShut {NoStop}%
\bibitem [{\citenamefont {Katano}\ \emph {et~al.}(1993)\citenamefont {Katano},
  \citenamefont {Funahashi}, \citenamefont {Môri}, \citenamefont {Ueda},\ and\
  \citenamefont {Fernandez-Baca}}]{Katano1993}%
  \BibitemOpen
  \bibfield  {author} {\bibinfo {author} {\bibfnamefont {Susumu}\ \bibnamefont
  {Katano}}, \bibinfo {author} {\bibfnamefont {Satoru}\ \bibnamefont
  {Funahashi}}, \bibinfo {author} {\bibfnamefont {Nobuo}\ \bibnamefont
  {Môri}}, \bibinfo {author} {\bibfnamefont {Yutaka}\ \bibnamefont {Ueda}}, \
  and\ \bibinfo {author} {\bibfnamefont {Jaime~A.}\ \bibnamefont
  {Fernandez-Baca}},\ }\bibfield  {title} {\enquote {\bibinfo {title} {Pressure
  effects on the structural phase transitions and superconductivity of
  {L}a$_{2-x}${B}a$_x${C}u{O}$_4$ ($x = 0.125$)},}\ }\href {\doibase
  10.1103/PhysRevB.48.6569} {\bibfield  {journal} {\bibinfo  {journal}
  {Physical Review B}\ }\textbf {\bibinfo {volume} {48}},\ \bibinfo {pages}
  {6569--6574} (\bibinfo {year} {1993})}\BibitemShut {NoStop}%
\bibitem [{\citenamefont {Arumugam}\ \emph
  {et~al.}(2002{\natexlab{a}})\citenamefont {Arumugam}, \citenamefont {Môri},
  \citenamefont {Takeshita}, \citenamefont {Takashima}, \citenamefont {Noda},
  \citenamefont {Eisaki},\ and\ \citenamefont {Uchida}}]{Arumugam2002}%
  \BibitemOpen
  \bibfield  {author} {\bibinfo {author} {\bibfnamefont {S.}~\bibnamefont
  {Arumugam}}, \bibinfo {author} {\bibfnamefont {N.}~\bibnamefont {Môri}},
  \bibinfo {author} {\bibfnamefont {N.}~\bibnamefont {Takeshita}}, \bibinfo
  {author} {\bibfnamefont {H.}~\bibnamefont {Takashima}}, \bibinfo {author}
  {\bibfnamefont {T.}~\bibnamefont {Noda}}, \bibinfo {author} {\bibfnamefont
  {H.}~\bibnamefont {Eisaki}}, \ and\ \bibinfo {author} {\bibfnamefont
  {S.}~\bibnamefont {Uchida}},\ }\bibfield  {title} {\enquote {\bibinfo {title}
  {Competition of {S}tatic {S}tripe and {S}uperconducting {P}hases in
  {L}a$_{1.48}${N}d$_{0.4}${S}r$_{0.12}${C}u{O}$_4$ {C}ontrolled by
  {P}ressure},}\ }\href {\doibase 10.1103/PhysRevLett.88.247001} {\bibfield
  {journal} {\bibinfo  {journal} {Physical Review Letters}\ }\textbf {\bibinfo
  {volume} {88}},\ \bibinfo {pages} {4} (\bibinfo {year}
  {2002}{\natexlab{a}})}\BibitemShut {NoStop}%
\bibitem [{\citenamefont {Arumugam}\ \emph
  {et~al.}(2002{\natexlab{b}})\citenamefont {Arumugam}, \citenamefont {Mydeen},
  \citenamefont {Manivannan}, \citenamefont {Mori}, \citenamefont {Ohashi},
  \citenamefont {Mori}, \citenamefont {Takeshita}, \citenamefont {Noda},
  \citenamefont {Eisaki},\ and\ \citenamefont {Uchida}}]{Arumugam2002a}%
  \BibitemOpen
  \bibfield  {author} {\bibinfo {author} {\bibfnamefont {S}~\bibnamefont
  {Arumugam}}, \bibinfo {author} {\bibfnamefont {K}~\bibnamefont {Mydeen}},
  \bibinfo {author} {\bibfnamefont {N}~\bibnamefont {Manivannan}}, \bibinfo
  {author} {\bibfnamefont {N}~\bibnamefont {Mori}}, \bibinfo {author}
  {\bibfnamefont {M}~\bibnamefont {Ohashi}}, \bibinfo {author} {\bibfnamefont
  {T}~\bibnamefont {Mori}}, \bibinfo {author} {\bibfnamefont {N}~\bibnamefont
  {Takeshita}}, \bibinfo {author} {\bibfnamefont {T}~\bibnamefont {Noda}},
  \bibinfo {author} {\bibfnamefont {H}~\bibnamefont {Eisaki}}, \ and\ \bibinfo
  {author} {\bibfnamefont {S}~\bibnamefont {Uchida}},\ }\bibfield  {title}
  {\enquote {\bibinfo {title} {Hydrostatic and uniaxial pressure effect on
  la1.45sr0.15nd0.4cuo4 single crystal},}\ }\href {\doibase
  https://doi.org/10.1016/S0921-4534(02)01409-0} {\bibfield  {journal}
  {\bibinfo  {journal} {Physica C: Superconductivity}\ }\textbf {\bibinfo
  {volume} {378-381}},\ \bibinfo {pages} {192--194} (\bibinfo {year}
  {2002}{\natexlab{b}})}\BibitemShut {NoStop}%
\bibitem [{\citenamefont {Takeshita}\ \emph {et~al.}(2004)\citenamefont
  {Takeshita}, \citenamefont {Sasagawa}, \citenamefont {Sugioka}, \citenamefont
  {Tokura},\ and\ \citenamefont {Takagi}}]{Takeshita2004}%
  \BibitemOpen
  \bibfield  {author} {\bibinfo {author} {\bibfnamefont {Nao}\ \bibnamefont
  {Takeshita}}, \bibinfo {author} {\bibfnamefont {Takao}\ \bibnamefont
  {Sasagawa}}, \bibinfo {author} {\bibfnamefont {Takenari}\ \bibnamefont
  {Sugioka}}, \bibinfo {author} {\bibfnamefont {Yoshinori}\ \bibnamefont
  {Tokura}}, \ and\ \bibinfo {author} {\bibfnamefont {Hidenori}\ \bibnamefont
  {Takagi}},\ }\bibfield  {title} {\enquote {\bibinfo {title} {Gigantic
  anisotropic uniaxial pressure effect on superconductivity within the
  {C}u{O}$_2$ plane of {L}a$_{1.64}${E}u$_{0.2}${S}r$_{0.16}${C}u{O}$_4$:
  {S}train control of stripe criticality},}\ }\href {\doibase
  10.1143/JPSJ.73.1123} {\bibfield  {journal} {\bibinfo  {journal} {Journal of
  the Physical Society of Japan}\ }\textbf {\bibinfo {volume} {73}},\ \bibinfo
  {pages} {1123--1126} (\bibinfo {year} {2004})}\BibitemShut {NoStop}%
\bibitem [{\citenamefont {Hücker}\ \emph {et~al.}(2010)\citenamefont
  {Hücker}, \citenamefont {Zimmermann}, \citenamefont {Debessai},
  \citenamefont {Schilling}, \citenamefont {Tranquada},\ and\ \citenamefont
  {Gu}}]{Hucker2010}%
  \BibitemOpen
  \bibfield  {author} {\bibinfo {author} {\bibfnamefont {M.}~\bibnamefont
  {Hücker}}, \bibinfo {author} {\bibfnamefont {M.~V.}\ \bibnamefont
  {Zimmermann}}, \bibinfo {author} {\bibfnamefont {M.}~\bibnamefont
  {Debessai}}, \bibinfo {author} {\bibfnamefont {J.~S.}\ \bibnamefont
  {Schilling}}, \bibinfo {author} {\bibfnamefont {J.~M.}\ \bibnamefont
  {Tranquada}}, \ and\ \bibinfo {author} {\bibfnamefont {G.~D.}\ \bibnamefont
  {Gu}},\ }\bibfield  {title} {\enquote {\bibinfo {title} {Spontaneous symmetry
  breaking by charge stripes in the high pressure phase of superconducting
  {L}a$_{1.875}${B}a$_{0.125}${C}u{O}$_4$},}\ }\href {\doibase
  10.1103/PhysRevLett.104.057004} {\bibfield  {journal} {\bibinfo  {journal}
  {Physical Review Letters}\ }\textbf {\bibinfo {volume} {104}},\ \bibinfo
  {pages} {057004} (\bibinfo {year} {2010})}\BibitemShut {NoStop}%
\bibitem [{\citenamefont {Guguchia}\ \emph {et~al.}(2020)\citenamefont
  {Guguchia}, \citenamefont {Das}, \citenamefont {Wang}, \citenamefont
  {Adachi}, \citenamefont {Kitajima}, \citenamefont {Elender}, \citenamefont
  {Brückner}, \citenamefont {Ghosh}, \citenamefont {Grinenko}, \citenamefont
  {Shiroka}, \citenamefont {Müller}, \citenamefont {Mudry}, \citenamefont
  {Baines}, \citenamefont {Bartkowiak}, \citenamefont {Koike}, \citenamefont
  {Amato}, \citenamefont {Tranquada}, \citenamefont {Klauss}, \citenamefont
  {Hicks},\ and\ \citenamefont {Luetkens}}]{Guguchia2020}%
  \BibitemOpen
  \bibfield  {author} {\bibinfo {author} {\bibfnamefont {Z.}~\bibnamefont
  {Guguchia}}, \bibinfo {author} {\bibfnamefont {D.}~\bibnamefont {Das}},
  \bibinfo {author} {\bibfnamefont {C.~N.}\ \bibnamefont {Wang}}, \bibinfo
  {author} {\bibfnamefont {T.}~\bibnamefont {Adachi}}, \bibinfo {author}
  {\bibfnamefont {N.}~\bibnamefont {Kitajima}}, \bibinfo {author}
  {\bibfnamefont {M.}~\bibnamefont {Elender}}, \bibinfo {author} {\bibfnamefont
  {F.}~\bibnamefont {Brückner}}, \bibinfo {author} {\bibfnamefont
  {S.}~\bibnamefont {Ghosh}}, \bibinfo {author} {\bibfnamefont
  {V.}~\bibnamefont {Grinenko}}, \bibinfo {author} {\bibfnamefont
  {T.}~\bibnamefont {Shiroka}}, \bibinfo {author} {\bibfnamefont
  {M.}~\bibnamefont {Müller}}, \bibinfo {author} {\bibfnamefont
  {C.}~\bibnamefont {Mudry}}, \bibinfo {author} {\bibfnamefont
  {C.}~\bibnamefont {Baines}}, \bibinfo {author} {\bibfnamefont
  {M.}~\bibnamefont {Bartkowiak}}, \bibinfo {author} {\bibfnamefont
  {Y.}~\bibnamefont {Koike}}, \bibinfo {author} {\bibfnamefont
  {A.}~\bibnamefont {Amato}}, \bibinfo {author} {\bibfnamefont {J.~M.}\
  \bibnamefont {Tranquada}}, \bibinfo {author} {\bibfnamefont {H.~H.}\
  \bibnamefont {Klauss}}, \bibinfo {author} {\bibfnamefont {C.~W.}\
  \bibnamefont {Hicks}}, \ and\ \bibinfo {author} {\bibfnamefont
  {H.}~\bibnamefont {Luetkens}},\ }\bibfield  {title} {\enquote {\bibinfo
  {title} {Using {U}niaxial {S}tress to {P}robe the {R}elationship between
  {C}ompeting {S}uperconducting {S}tates in a {C}uprate with {S}pin-stripe
  {O}rder},}\ }\href {\doibase 10.1103/PhysRevLett.125.097005} {\bibfield
  {journal} {\bibinfo  {journal} {Physical Review Letters}\ }\textbf {\bibinfo
  {volume} {125}},\ \bibinfo {pages} {97005} (\bibinfo {year}
  {2020})}\BibitemShut {NoStop}%
\bibitem [{\citenamefont {Boyle}\ \emph {et~al.}(2021)\citenamefont {Boyle},
  \citenamefont {Walker}, \citenamefont {Ruiz}, \citenamefont {Schierle},
  \citenamefont {Zhao}, \citenamefont {Boschini}, \citenamefont {Sutarto},
  \citenamefont {Boyko}, \citenamefont {Moore}, \citenamefont {Tamura},
  \citenamefont {He}, \citenamefont {Weschke}, \citenamefont {Gozar},
  \citenamefont {Peng}, \citenamefont {Komarek}, \citenamefont {Damascelli},
  \citenamefont {Schüßler-Langeheine}, \citenamefont {Frano}, \citenamefont
  {Neto},\ and\ \citenamefont {Blanco-Canosa}}]{Boyle2021}%
  \BibitemOpen
  \bibfield  {author} {\bibinfo {author} {\bibfnamefont {T.~J.}\ \bibnamefont
  {Boyle}}, \bibinfo {author} {\bibfnamefont {M.}~\bibnamefont {Walker}},
  \bibinfo {author} {\bibfnamefont {A.}~\bibnamefont {Ruiz}}, \bibinfo {author}
  {\bibfnamefont {E.}~\bibnamefont {Schierle}}, \bibinfo {author}
  {\bibfnamefont {Z.}~\bibnamefont {Zhao}}, \bibinfo {author} {\bibfnamefont
  {F.}~\bibnamefont {Boschini}}, \bibinfo {author} {\bibfnamefont
  {R.}~\bibnamefont {Sutarto}}, \bibinfo {author} {\bibfnamefont {T.~D.}\
  \bibnamefont {Boyko}}, \bibinfo {author} {\bibfnamefont {W.}~\bibnamefont
  {Moore}}, \bibinfo {author} {\bibfnamefont {N.}~\bibnamefont {Tamura}},
  \bibinfo {author} {\bibfnamefont {F.}~\bibnamefont {He}}, \bibinfo {author}
  {\bibfnamefont {E.}~\bibnamefont {Weschke}}, \bibinfo {author} {\bibfnamefont
  {A.}~\bibnamefont {Gozar}}, \bibinfo {author} {\bibfnamefont
  {W.}~\bibnamefont {Peng}}, \bibinfo {author} {\bibfnamefont {A.~C.}\
  \bibnamefont {Komarek}}, \bibinfo {author} {\bibfnamefont {A.}~\bibnamefont
  {Damascelli}}, \bibinfo {author} {\bibfnamefont {C.}~\bibnamefont
  {Schüßler-Langeheine}}, \bibinfo {author} {\bibfnamefont {A.}~\bibnamefont
  {Frano}}, \bibinfo {author} {\bibfnamefont {E.~H. Da~Silva}\ \bibnamefont
  {Neto}}, \ and\ \bibinfo {author} {\bibfnamefont {S.}~\bibnamefont
  {Blanco-Canosa}},\ }\bibfield  {title} {\enquote {\bibinfo {title} {Large
  response of charge stripes to uniaxial stress in
  {L}a$_{1.475}${N}d$_{0.4}${S}r$_{0.125}${C}u{O}$_4$},}\ }\href {\doibase
  10.1103/PhysRevResearch.3.L022004} {\bibfield  {journal} {\bibinfo  {journal}
  {Physical Review Research}\ }\textbf {\bibinfo {volume} {3}},\ \bibinfo
  {pages} {1--6} (\bibinfo {year} {2021})}\BibitemShut {NoStop}%
\bibitem [{\citenamefont {Guguchia}\ \emph {et~al.}(2023)\citenamefont
  {Guguchia}, \citenamefont {Das}, \citenamefont {Simutis}, \citenamefont
  {Adachi}, \citenamefont {Küspert}, \citenamefont {Kitajima}, \citenamefont
  {Elender}, \citenamefont {Grinenko}, \citenamefont {Ivashko}, \citenamefont
  {Zimmermann}, \citenamefont {Müller}, \citenamefont {Mielke}, \citenamefont
  {Hotz}, \citenamefont {Mudry}, \citenamefont {Baines}, \citenamefont
  {Bartkowiak}, \citenamefont {Shiroka}, \citenamefont {Koike}, \citenamefont
  {Amato}, \citenamefont {Hicks}, \citenamefont {Gu}, \citenamefont
  {Tranquada}, \citenamefont {Klauss}, \citenamefont {Chang}, \citenamefont
  {Janoschek},\ and\ \citenamefont {Luetkens}}]{Guguchia2023}%
  \BibitemOpen
  \bibfield  {author} {\bibinfo {author} {\bibfnamefont {Z.}~\bibnamefont
  {Guguchia}}, \bibinfo {author} {\bibfnamefont {D.}~\bibnamefont {Das}},
  \bibinfo {author} {\bibfnamefont {G.}~\bibnamefont {Simutis}}, \bibinfo
  {author} {\bibfnamefont {T.}~\bibnamefont {Adachi}}, \bibinfo {author}
  {\bibfnamefont {J.}~\bibnamefont {Küspert}}, \bibinfo {author}
  {\bibfnamefont {N.}~\bibnamefont {Kitajima}}, \bibinfo {author}
  {\bibfnamefont {M.}~\bibnamefont {Elender}}, \bibinfo {author} {\bibfnamefont
  {V.}~\bibnamefont {Grinenko}}, \bibinfo {author} {\bibfnamefont
  {O.}~\bibnamefont {Ivashko}}, \bibinfo {author} {\bibfnamefont {M.~v.}\
  \bibnamefont {Zimmermann}}, \bibinfo {author} {\bibfnamefont
  {M.}~\bibnamefont {Müller}}, \bibinfo {author} {\bibfnamefont
  {C.}~\bibnamefont {Mielke}}, \bibinfo {author} {\bibfnamefont
  {F.}~\bibnamefont {Hotz}}, \bibinfo {author} {\bibfnamefont {C.}~\bibnamefont
  {Mudry}}, \bibinfo {author} {\bibfnamefont {C.}~\bibnamefont {Baines}},
  \bibinfo {author} {\bibfnamefont {M.}~\bibnamefont {Bartkowiak}}, \bibinfo
  {author} {\bibfnamefont {T.}~\bibnamefont {Shiroka}}, \bibinfo {author}
  {\bibfnamefont {Y.}~\bibnamefont {Koike}}, \bibinfo {author} {\bibfnamefont
  {A.}~\bibnamefont {Amato}}, \bibinfo {author} {\bibfnamefont {C.~W.}\
  \bibnamefont {Hicks}}, \bibinfo {author} {\bibfnamefont {G.~D.}\ \bibnamefont
  {Gu}}, \bibinfo {author} {\bibfnamefont {J.~M.}\ \bibnamefont {Tranquada}},
  \bibinfo {author} {\bibfnamefont {H.~H.}\ \bibnamefont {Klauss}}, \bibinfo
  {author} {\bibfnamefont {J.~J.}\ \bibnamefont {Chang}}, \bibinfo {author}
  {\bibfnamefont {M.}~\bibnamefont {Janoschek}}, \ and\ \bibinfo {author}
  {\bibfnamefont {H.}~\bibnamefont {Luetkens}},\ }\href {\doibase
  10.48550/ARXIV.2302.07015} {\enquote {\bibinfo {title} {Designing the
  stripe-ordered cuprate phase diagram through uniaxial-stress},}\ } (\bibinfo
  {year} {2023})\BibitemShut {NoStop}%
\bibitem [{\citenamefont {Tranquada}(2020)}]{Tranquada2020}%
  \BibitemOpen
  \bibfield  {author} {\bibinfo {author} {\bibfnamefont {J.~M.}\ \bibnamefont
  {Tranquada}},\ }\bibfield  {title} {\enquote {\bibinfo {title} {Cuprate
  superconductors as viewed through a striped lens},}\ }\href {\doibase
  10.1080/00018732.2021.1935698} {\bibfield  {journal} {\bibinfo  {journal}
  {Advances in Physics}\ }\textbf {\bibinfo {volume} {69}},\ \bibinfo {pages}
  {437--509} (\bibinfo {year} {2020})}\BibitemShut {NoStop}%
\bibitem [{\citenamefont {Tidey}\ \emph
  {et~al.}(2022{\natexlab{a}})\citenamefont {Tidey}, \citenamefont {Keegan},
  \citenamefont {Bristowe}, \citenamefont {Mostofi}, \citenamefont {Hong},
  \citenamefont {Chen}, \citenamefont {Chuang}, \citenamefont {Chen},\ and\
  \citenamefont {Senn}}]{Tidey2022}%
  \BibitemOpen
  \bibfield  {author} {\bibinfo {author} {\bibfnamefont {Jeremiah~P.}\
  \bibnamefont {Tidey}}, \bibinfo {author} {\bibfnamefont {Christopher}\
  \bibnamefont {Keegan}}, \bibinfo {author} {\bibfnamefont {Nicholas~C.}\
  \bibnamefont {Bristowe}}, \bibinfo {author} {\bibfnamefont {Arash~A.}\
  \bibnamefont {Mostofi}}, \bibinfo {author} {\bibfnamefont {Zih-Mei}\
  \bibnamefont {Hong}}, \bibinfo {author} {\bibfnamefont {Bo-Hao}\ \bibnamefont
  {Chen}}, \bibinfo {author} {\bibfnamefont {Yu-Chun}\ \bibnamefont {Chuang}},
  \bibinfo {author} {\bibfnamefont {Wei-Tin}\ \bibnamefont {Chen}}, \ and\
  \bibinfo {author} {\bibfnamefont {Mark~S.}\ \bibnamefont {Senn}},\ }\bibfield
   {title} {\enquote {\bibinfo {title} {Structural origins of the
  low-temperature orthorhombic to low-temperature tetragonal phase transition
  in high-{T}$_\mathrm{c}$ cuprates},}\ }\href {\doibase
  10.1103/PhysRevB.106.085112} {\bibfield  {journal} {\bibinfo  {journal}
  {Physical Review B}\ }\textbf {\bibinfo {volume} {106}},\ \bibinfo {pages}
  {1--6} (\bibinfo {year} {2022}{\natexlab{a}})}\BibitemShut {NoStop}%
\bibitem [{\citenamefont {Shannon}(1976)}]{Shannon1976}%
  \BibitemOpen
  \bibfield  {author} {\bibinfo {author} {\bibfnamefont {R.~D.}\ \bibnamefont
  {Shannon}},\ }\bibfield  {title} {\enquote {\bibinfo {title} {Revised
  effective ionic radii and systematic studies of interatomic distances in
  halides and chalcogenides},}\ }\href {\doibase 10.1107/S0567739476001551}
  {\bibfield  {journal} {\bibinfo  {journal} {Acta Crystallographica Section
  A}\ }\textbf {\bibinfo {volume} {32}},\ \bibinfo {pages} {751--767} (\bibinfo
  {year} {1976})}\BibitemShut {NoStop}%
\bibitem [{\citenamefont {Giannozzi}\ \emph {et~al.}(2009)\citenamefont
  {Giannozzi}, \citenamefont {Baroni}, \citenamefont {Bonini}, \citenamefont
  {Calandra}, \citenamefont {Car}, \citenamefont {Cavazzoni}, \citenamefont
  {Ceresoli}, \citenamefont {Chiarotti}, \citenamefont {Cococcioni},
  \citenamefont {Dabo}, \citenamefont {Corso}, \citenamefont {de~Gironcoli},
  \citenamefont {Fabris}, \citenamefont {Fratesi}, \citenamefont {Gebauer},
  \citenamefont {Gerstmann}, \citenamefont {Gougoussis}, \citenamefont
  {Kokalj}, \citenamefont {Lazzeri}, \citenamefont {Martin-Samos},
  \citenamefont {Marzari}, \citenamefont {Mauri}, \citenamefont {Mazzarello},
  \citenamefont {Paolini}, \citenamefont {Pasquarello}, \citenamefont
  {Paulatto}, \citenamefont {Sbraccia}, \citenamefont {Scandolo}, \citenamefont
  {Sclauzero}, \citenamefont {Seitsonen}, \citenamefont {Smogunov},
  \citenamefont {Umari},\ and\ \citenamefont {Wentzcovitch}}]{QE-2009}%
  \BibitemOpen
  \bibfield  {author} {\bibinfo {author} {\bibfnamefont {P.}~\bibnamefont
  {Giannozzi}}, \bibinfo {author} {\bibfnamefont {S.}~\bibnamefont {Baroni}},
  \bibinfo {author} {\bibfnamefont {N.}~\bibnamefont {Bonini}}, \bibinfo
  {author} {\bibfnamefont {M.}~\bibnamefont {Calandra}}, \bibinfo {author}
  {\bibfnamefont {R.}~\bibnamefont {Car}}, \bibinfo {author} {\bibfnamefont
  {C.}~\bibnamefont {Cavazzoni}}, \bibinfo {author} {\bibfnamefont
  {D.}~\bibnamefont {Ceresoli}}, \bibinfo {author} {\bibfnamefont {G.~L.}\
  \bibnamefont {Chiarotti}}, \bibinfo {author} {\bibfnamefont {M.}~\bibnamefont
  {Cococcioni}}, \bibinfo {author} {\bibfnamefont {I.}~\bibnamefont {Dabo}},
  \bibinfo {author} {\bibfnamefont {A.~Dal}\ \bibnamefont {Corso}}, \bibinfo
  {author} {\bibfnamefont {S.}~\bibnamefont {de~Gironcoli}}, \bibinfo {author}
  {\bibfnamefont {S.}~\bibnamefont {Fabris}}, \bibinfo {author} {\bibfnamefont
  {G.}~\bibnamefont {Fratesi}}, \bibinfo {author} {\bibfnamefont
  {R.}~\bibnamefont {Gebauer}}, \bibinfo {author} {\bibfnamefont
  {U.}~\bibnamefont {Gerstmann}}, \bibinfo {author} {\bibfnamefont
  {C.}~\bibnamefont {Gougoussis}}, \bibinfo {author} {\bibfnamefont
  {A.}~\bibnamefont {Kokalj}}, \bibinfo {author} {\bibfnamefont
  {M.}~\bibnamefont {Lazzeri}}, \bibinfo {author} {\bibfnamefont
  {L.}~\bibnamefont {Martin-Samos}}, \bibinfo {author} {\bibfnamefont
  {N.}~\bibnamefont {Marzari}}, \bibinfo {author} {\bibfnamefont
  {F.}~\bibnamefont {Mauri}}, \bibinfo {author} {\bibfnamefont
  {R.}~\bibnamefont {Mazzarello}}, \bibinfo {author} {\bibfnamefont
  {S.}~\bibnamefont {Paolini}}, \bibinfo {author} {\bibfnamefont
  {A.}~\bibnamefont {Pasquarello}}, \bibinfo {author} {\bibfnamefont
  {L.}~\bibnamefont {Paulatto}}, \bibinfo {author} {\bibfnamefont
  {C.}~\bibnamefont {Sbraccia}}, \bibinfo {author} {\bibfnamefont
  {S.}~\bibnamefont {Scandolo}}, \bibinfo {author} {\bibfnamefont
  {G.}~\bibnamefont {Sclauzero}}, \bibinfo {author} {\bibfnamefont {A.~P.}\
  \bibnamefont {Seitsonen}}, \bibinfo {author} {\bibfnamefont {A.}~\bibnamefont
  {Smogunov}}, \bibinfo {author} {\bibfnamefont {P.}~\bibnamefont {Umari}}, \
  and\ \bibinfo {author} {\bibfnamefont {R.~M.}\ \bibnamefont {Wentzcovitch}},\
  }\bibfield  {title} {\enquote {\bibinfo {title} {Quantum espresso: a modular
  and open-source software project for quantum simulations of materials},}\
  }\href {http://www.quantum-espresso.org} {\bibfield  {journal} {\bibinfo
  {journal} {Journal of Physics: Condensed Matter}\ }\textbf {\bibinfo {volume}
  {21}},\ \bibinfo {pages} {395502 (19pp)} (\bibinfo {year}
  {2009})}\BibitemShut {NoStop}%
\bibitem [{\citenamefont {Giannozzi}\ \emph {et~al.}(2017)\citenamefont
  {Giannozzi}, \citenamefont {Andreussi}, \citenamefont {Brumme}, \citenamefont
  {Bunau}, \citenamefont {Nardelli}, \citenamefont {Calandra}, \citenamefont
  {Car}, \citenamefont {Cavazzoni}, \citenamefont {Ceresoli}, \citenamefont
  {Cococcioni}, \citenamefont {Colonna}, \citenamefont {Carnimeo},
  \citenamefont {Corso}, \citenamefont {de~Gironcoli}, \citenamefont {Delugas},
  \citenamefont {DiStasio}, \citenamefont {Ferretti}, \citenamefont {Floris},
  \citenamefont {Fratesi}, \citenamefont {Fugallo}, \citenamefont {Gebauer},
  \citenamefont {Gerstmann}, \citenamefont {Giustino}, \citenamefont {Gorni},
  \citenamefont {Jia}, \citenamefont {Kawamura}, \citenamefont {Ko},
  \citenamefont {Kokalj}, \citenamefont {Küçükbenli}, \citenamefont
  {Lazzeri}, \citenamefont {Marsili}, \citenamefont {Marzari}, \citenamefont
  {Mauri}, \citenamefont {Nguyen}, \citenamefont {Nguyen}, \citenamefont {de-la
  Roza}, \citenamefont {Paulatto}, \citenamefont {Poncé}, \citenamefont
  {Rocca}, \citenamefont {Sabatini}, \citenamefont {Santra}, \citenamefont
  {Schlipf}, \citenamefont {Seitsonen}, \citenamefont {Smogunov}, \citenamefont
  {Timrov}, \citenamefont {Thonhauser}, \citenamefont {Umari}, \citenamefont
  {Vast}, \citenamefont {Wu},\ and\ \citenamefont {Baroni}}]{QE-2017}%
  \BibitemOpen
  \bibfield  {author} {\bibinfo {author} {\bibfnamefont {P.}~\bibnamefont
  {Giannozzi}}, \bibinfo {author} {\bibfnamefont {O.}~\bibnamefont
  {Andreussi}}, \bibinfo {author} {\bibfnamefont {T.}~\bibnamefont {Brumme}},
  \bibinfo {author} {\bibfnamefont {O.}~\bibnamefont {Bunau}}, \bibinfo
  {author} {\bibfnamefont {M.~B.}\ \bibnamefont {Nardelli}}, \bibinfo {author}
  {\bibfnamefont {M.}~\bibnamefont {Calandra}}, \bibinfo {author}
  {\bibfnamefont {R.}~\bibnamefont {Car}}, \bibinfo {author} {\bibfnamefont
  {C.}~\bibnamefont {Cavazzoni}}, \bibinfo {author} {\bibfnamefont
  {D}~\bibnamefont {Ceresoli}}, \bibinfo {author} {\bibfnamefont
  {M}~\bibnamefont {Cococcioni}}, \bibinfo {author} {\bibfnamefont
  {N}~\bibnamefont {Colonna}}, \bibinfo {author} {\bibfnamefont
  {I.}~\bibnamefont {Carnimeo}}, \bibinfo {author} {\bibfnamefont {A.~D.}\
  \bibnamefont {Corso}}, \bibinfo {author} {\bibfnamefont {S.}~\bibnamefont
  {de~Gironcoli}}, \bibinfo {author} {\bibfnamefont {P.}~\bibnamefont
  {Delugas}}, \bibinfo {author} {\bibfnamefont {R.~A.~Jr.}\ \bibnamefont
  {DiStasio}}, \bibinfo {author} {\bibfnamefont {A.}~\bibnamefont {Ferretti}},
  \bibinfo {author} {\bibfnamefont {A.}~\bibnamefont {Floris}}, \bibinfo
  {author} {\bibfnamefont {G.}~\bibnamefont {Fratesi}}, \bibinfo {author}
  {\bibfnamefont {G.}~\bibnamefont {Fugallo}}, \bibinfo {author} {\bibfnamefont
  {R.}~\bibnamefont {Gebauer}}, \bibinfo {author} {\bibfnamefont
  {U.}~\bibnamefont {Gerstmann}}, \bibinfo {author} {\bibfnamefont
  {F.}~\bibnamefont {Giustino}}, \bibinfo {author} {\bibfnamefont
  {T.}~\bibnamefont {Gorni}}, \bibinfo {author} {\bibfnamefont
  {J.}~\bibnamefont {Jia}}, \bibinfo {author} {\bibfnamefont {M.}~\bibnamefont
  {Kawamura}}, \bibinfo {author} {\bibfnamefont {H.-Y.}\ \bibnamefont {Ko}},
  \bibinfo {author} {\bibfnamefont {A.}~\bibnamefont {Kokalj}}, \bibinfo
  {author} {\bibfnamefont {E.}~\bibnamefont {Küçükbenli}}, \bibinfo {author}
  {\bibfnamefont {M.}~\bibnamefont {Lazzeri}}, \bibinfo {author} {\bibfnamefont
  {M.}~\bibnamefont {Marsili}}, \bibinfo {author} {\bibfnamefont
  {N.}~\bibnamefont {Marzari}}, \bibinfo {author} {\bibfnamefont
  {F.}~\bibnamefont {Mauri}}, \bibinfo {author} {\bibfnamefont {N.~L.}\
  \bibnamefont {Nguyen}}, \bibinfo {author} {\bibfnamefont {H.-V.}\
  \bibnamefont {Nguyen}}, \bibinfo {author} {\bibfnamefont {A.~Otero}\
  \bibnamefont {de-la Roza}}, \bibinfo {author} {\bibfnamefont
  {L.}~\bibnamefont {Paulatto}}, \bibinfo {author} {\bibfnamefont
  {S.}~\bibnamefont {Poncé}}, \bibinfo {author} {\bibfnamefont
  {D.}~\bibnamefont {Rocca}}, \bibinfo {author} {\bibfnamefont
  {R.}~\bibnamefont {Sabatini}}, \bibinfo {author} {\bibfnamefont
  {B.}~\bibnamefont {Santra}}, \bibinfo {author} {\bibfnamefont
  {M.}~\bibnamefont {Schlipf}}, \bibinfo {author} {\bibfnamefont {A.~P.}\
  \bibnamefont {Seitsonen}}, \bibinfo {author} {\bibfnamefont {A.}~\bibnamefont
  {Smogunov}}, \bibinfo {author} {\bibfnamefont {I.}~\bibnamefont {Timrov}},
  \bibinfo {author} {\bibfnamefont {T.}~\bibnamefont {Thonhauser}}, \bibinfo
  {author} {\bibfnamefont {P.}~\bibnamefont {Umari}}, \bibinfo {author}
  {\bibfnamefont {N.}~\bibnamefont {Vast}}, \bibinfo {author} {\bibfnamefont
  {X.}~\bibnamefont {Wu}}, \ and\ \bibinfo {author} {\bibfnamefont
  {S.}~\bibnamefont {Baroni}},\ }\bibfield  {title} {\enquote {\bibinfo {title}
  {Advanced capabilities for materials modelling with quantum espresso},}\
  }\href {http://stacks.iop.org/0953-8984/29/i=46/a=465901} {\bibfield
  {journal} {\bibinfo  {journal} {Journal of Physics: Condensed Matter}\
  }\textbf {\bibinfo {volume} {29}},\ \bibinfo {pages} {465901} (\bibinfo
  {year} {2017})}\BibitemShut {NoStop}%
\bibitem [{\citenamefont {Perdew}\ \emph {et~al.}(2008)\citenamefont {Perdew},
  \citenamefont {Ruzsinszky}, \citenamefont {Csonka}, \citenamefont {Vydrov},
  \citenamefont {Scuseria}, \citenamefont {Constantin}, \citenamefont {Zhou},\
  and\ \citenamefont {Burke}}]{Perdew2008}%
  \BibitemOpen
  \bibfield  {author} {\bibinfo {author} {\bibfnamefont {J.~P.}\ \bibnamefont
  {Perdew}}, \bibinfo {author} {\bibfnamefont {A.}~\bibnamefont {Ruzsinszky}},
  \bibinfo {author} {\bibfnamefont {G.~I.}\ \bibnamefont {Csonka}}, \bibinfo
  {author} {\bibfnamefont {O.~A.}\ \bibnamefont {Vydrov}}, \bibinfo {author}
  {\bibfnamefont {G.~E.}\ \bibnamefont {Scuseria}}, \bibinfo {author}
  {\bibfnamefont {L.~A.}\ \bibnamefont {Constantin}}, \bibinfo {author}
  {\bibfnamefont {X.}~\bibnamefont {Zhou}}, \ and\ \bibinfo {author}
  {\bibfnamefont {K.}~\bibnamefont {Burke}},\ }\bibfield  {title} {\enquote
  {\bibinfo {title} {Restoring the density-gradient expansion for exchange in
  solids and surfaces},}\ }\href {\doibase 10.1103/PhysRevLett.100.136406}
  {\bibfield  {journal} {\bibinfo  {journal} {Physical Review Letters}\
  }\textbf {\bibinfo {volume} {100}},\ \bibinfo {pages} {136406} (\bibinfo
  {year} {2008})}\BibitemShut {NoStop}%
\bibitem [{\citenamefont {Garrity}\ \emph {et~al.}(2014)\citenamefont
  {Garrity}, \citenamefont {Bennett}, \citenamefont {Rabe},\ and\ \citenamefont
  {Vanderbilt}}]{Garrity2014}%
  \BibitemOpen
  \bibfield  {author} {\bibinfo {author} {\bibfnamefont {Kevin~F.}\
  \bibnamefont {Garrity}}, \bibinfo {author} {\bibfnamefont {Joseph~W.}\
  \bibnamefont {Bennett}}, \bibinfo {author} {\bibfnamefont {Karin~M.}\
  \bibnamefont {Rabe}}, \ and\ \bibinfo {author} {\bibfnamefont {David}\
  \bibnamefont {Vanderbilt}},\ }\bibfield  {title} {\enquote {\bibinfo {title}
  {Pseudopotentials for high-throughput dft calculations},}\ }\href {\doibase
  10.1016/j.commatsci.2013.08.053} {\bibfield  {journal} {\bibinfo  {journal}
  {Computational Materials Science}\ }\textbf {\bibinfo {volume} {81}},\
  \bibinfo {pages} {446--452} (\bibinfo {year} {2014})}\BibitemShut {NoStop}%
\bibitem [{\citenamefont {Keegan}\ \emph {et~al.}(2022)\citenamefont {Keegan},
  \citenamefont {Senn}, \citenamefont {Bristowe},\ and\ \citenamefont
  {Mostofi}}]{SI_LBO2022}%
  \BibitemOpen
  \bibfield  {author} {\bibinfo {author} {\bibfnamefont {C.}~\bibnamefont
  {Keegan}}, \bibinfo {author} {\bibfnamefont {M.~S.}\ \bibnamefont {Senn}},
  \bibinfo {author} {\bibfnamefont {N.~C.}\ \bibnamefont {Bristowe}}, \ and\
  \bibinfo {author} {\bibfnamefont {A.~A.}\ \bibnamefont {Mostofi}},\
  }\href@noop {} {\enquote {\bibinfo {title} {Supplementary {I}nformation},}\ }
  (\bibinfo {year} {2022})\BibitemShut {NoStop}%
\bibitem [{\citenamefont {Stokes}\ and\ \citenamefont
  {Hatch}(2005)}]{FINDSYM_Stokes2005}%
  \BibitemOpen
  \bibfield  {author} {\bibinfo {author} {\bibfnamefont {Harold~T}\
  \bibnamefont {Stokes}}\ and\ \bibinfo {author} {\bibfnamefont {Dorian~M}\
  \bibnamefont {Hatch}},\ }\bibfield  {title} {\enquote {\bibinfo {title}
  {Findsym: program for identifying the space-group symmetry of a crystal},}\
  }\href {\doibase 10.1107/S0021889804031528} {\bibfield  {journal} {\bibinfo
  {journal} {J. Appl. Cryst}\ }\textbf {\bibinfo {volume} {38}},\ \bibinfo
  {pages} {237--238} (\bibinfo {year} {2005})}\BibitemShut {NoStop}%
\bibitem [{\citenamefont {Campbell}\ \emph {et~al.}(2006)\citenamefont
  {Campbell}, \citenamefont {Stokes}, \citenamefont {Tanner},\ and\
  \citenamefont {Hatch}}]{ISODISPLACE_Campbell2006}%
  \BibitemOpen
  \bibfield  {author} {\bibinfo {author} {\bibfnamefont {B.~J.}\ \bibnamefont
  {Campbell}}, \bibinfo {author} {\bibfnamefont {H.~T.}\ \bibnamefont
  {Stokes}}, \bibinfo {author} {\bibfnamefont {D.~E.}\ \bibnamefont {Tanner}},
  \ and\ \bibinfo {author} {\bibfnamefont {D.~M.}\ \bibnamefont {Hatch}},\
  }\bibfield  {title} {\enquote {\bibinfo {title} {{ISODISPLACE} : a web-based
  tool for exploring structural distortions},}\ }\href {\doibase
  10.1107/S0021889806014075} {\bibfield  {journal} {\bibinfo  {journal} {J.
  Appl. Cryst}\ }\textbf {\bibinfo {volume} {39}},\ \bibinfo {pages} {607--614}
  (\bibinfo {year} {2006})}\BibitemShut {NoStop}%
\bibitem [{\citenamefont {Momma}\ and\ \citenamefont
  {Izumi}(2011)}]{VESTA_2011}%
  \BibitemOpen
  \bibfield  {author} {\bibinfo {author} {\bibfnamefont {Koichi}\ \bibnamefont
  {Momma}}\ and\ \bibinfo {author} {\bibfnamefont {Fujio}\ \bibnamefont
  {Izumi}},\ }\bibfield  {title} {\enquote {\bibinfo {title} {{{\it VESTA3} for
  three-dimensional visualization of crystal, volumetric and morphology
  data}},}\ }\href {\doibase 10.1107/S0021889811038970} {\bibfield  {journal}
  {\bibinfo  {journal} {Journal of Applied Crystallography}\ }\textbf {\bibinfo
  {volume} {44}},\ \bibinfo {pages} {1272--1276} (\bibinfo {year}
  {2011})}\BibitemShut {NoStop}%
\bibitem [{\citenamefont {Radaelli}\ \emph {et~al.}(1994)\citenamefont
  {Radaelli}, \citenamefont {Hinks}, \citenamefont {Mitchell}, \citenamefont
  {Hunter}, \citenamefont {Wagner}, \citenamefont {Dabrowski}, \citenamefont
  {Vandervoort}, \citenamefont {Viswanathan},\ and\ \citenamefont
  {Jorgensen}}]{Radaelli1994}%
  \BibitemOpen
  \bibfield  {author} {\bibinfo {author} {\bibfnamefont {P.~G.}\ \bibnamefont
  {Radaelli}}, \bibinfo {author} {\bibfnamefont {D.~G.}\ \bibnamefont {Hinks}},
  \bibinfo {author} {\bibfnamefont {A.~W.}\ \bibnamefont {Mitchell}}, \bibinfo
  {author} {\bibfnamefont {B.~A.}\ \bibnamefont {Hunter}}, \bibinfo {author}
  {\bibfnamefont {J.~L.}\ \bibnamefont {Wagner}}, \bibinfo {author}
  {\bibfnamefont {B.}~\bibnamefont {Dabrowski}}, \bibinfo {author}
  {\bibfnamefont {K.~G.}\ \bibnamefont {Vandervoort}}, \bibinfo {author}
  {\bibfnamefont {H.~K.}\ \bibnamefont {Viswanathan}}, \ and\ \bibinfo {author}
  {\bibfnamefont {J.~D.}\ \bibnamefont {Jorgensen}},\ }\bibfield  {title}
  {\enquote {\bibinfo {title} {Structural and superconducting properties of
  {L}a$_{2-x}${S}r$_x${C}u{O}$_4$ as a function of {S}r content},}\ }\href
  {\doibase 10.1103/PhysRevB.49.4163} {\bibfield  {journal} {\bibinfo
  {journal} {Physical Review B}\ }\textbf {\bibinfo {volume} {49}},\ \bibinfo
  {pages} {4163--4175} (\bibinfo {year} {1994})}\BibitemShut {NoStop}%
\bibitem [{\citenamefont {Zhang}\ \emph {et~al.}(1992)\citenamefont {Zhang},
  \citenamefont {Catlow}, \citenamefont {Parker},\ and\ \citenamefont
  {Wall}}]{Zhang1992}%
  \BibitemOpen
  \bibfield  {author} {\bibinfo {author} {\bibfnamefont {X.}~\bibnamefont
  {Zhang}}, \bibinfo {author} {\bibfnamefont {C.R.A.}\ \bibnamefont {Catlow}},
  \bibinfo {author} {\bibfnamefont {S.C.}\ \bibnamefont {Parker}}, \ and\
  \bibinfo {author} {\bibfnamefont {A.}~\bibnamefont {Wall}},\ }\bibfield
  {title} {\enquote {\bibinfo {title} {Simulation study of pressure-induced
  structural changes in la2cuo4 and in la1.83sr0.17cuo4},}\ }\href {\doibase
  https://doi.org/10.1016/0022-3697(92)90186-H} {\bibfield  {journal} {\bibinfo
   {journal} {Journal of Physics and Chemistry of Solids}\ }\textbf {\bibinfo
  {volume} {53}},\ \bibinfo {pages} {761--770} (\bibinfo {year}
  {1992})}\BibitemShut {NoStop}%
\bibitem [{\citenamefont {Furness}\ \emph {et~al.}(2018)\citenamefont
  {Furness}, \citenamefont {Zhang}, \citenamefont {Lane}, \citenamefont {Buda},
  \citenamefont {Barbiellini}, \citenamefont {Markiewicz}, \citenamefont
  {Bansil},\ and\ \citenamefont {Sun}}]{Furness2018}%
  \BibitemOpen
  \bibfield  {author} {\bibinfo {author} {\bibfnamefont {J.~W.}\ \bibnamefont
  {Furness}}, \bibinfo {author} {\bibfnamefont {Y.}~\bibnamefont {Zhang}},
  \bibinfo {author} {\bibfnamefont {C.}~\bibnamefont {Lane}}, \bibinfo {author}
  {\bibfnamefont {I.~G.}\ \bibnamefont {Buda}}, \bibinfo {author}
  {\bibfnamefont {B.}~\bibnamefont {Barbiellini}}, \bibinfo {author}
  {\bibfnamefont {R.~S.}\ \bibnamefont {Markiewicz}}, \bibinfo {author}
  {\bibfnamefont {A.}~\bibnamefont {Bansil}}, \ and\ \bibinfo {author}
  {\bibfnamefont {J.}~\bibnamefont {Sun}},\ }\bibfield  {title} {\enquote
  {\bibinfo {title} {An accurate first-principles treatment of doping-dependent
  electronic structure of high-temperature cuprate superconductors},}\ }\href
  {\doibase 10.1038/s42005-018-0009-4} {\bibfield  {journal} {\bibinfo
  {journal} {Communications Physics}\ }\textbf {\bibinfo {volume} {1}},\
  \bibinfo {pages} {11} (\bibinfo {year} {2018})}\BibitemShut {NoStop}%
\bibitem [{\citenamefont {Pokharel}\ \emph {et~al.}(2022)\citenamefont
  {Pokharel}, \citenamefont {Lane}, \citenamefont {Furness}, \citenamefont
  {Zhang}, \citenamefont {Ning}, \citenamefont {Barbiellini}, \citenamefont
  {Markiewicz}, \citenamefont {Zhang}, \citenamefont {Bansil},\ and\
  \citenamefont {Sun}}]{Pokharel2022}%
  \BibitemOpen
  \bibfield  {author} {\bibinfo {author} {\bibfnamefont {Kanun}\ \bibnamefont
  {Pokharel}}, \bibinfo {author} {\bibfnamefont {Christopher}\ \bibnamefont
  {Lane}}, \bibinfo {author} {\bibfnamefont {James~W.}\ \bibnamefont
  {Furness}}, \bibinfo {author} {\bibfnamefont {Ruiqi}\ \bibnamefont {Zhang}},
  \bibinfo {author} {\bibfnamefont {Jinliang}\ \bibnamefont {Ning}}, \bibinfo
  {author} {\bibfnamefont {Bernardo}\ \bibnamefont {Barbiellini}}, \bibinfo
  {author} {\bibfnamefont {Robert~S.}\ \bibnamefont {Markiewicz}}, \bibinfo
  {author} {\bibfnamefont {Yubo}\ \bibnamefont {Zhang}}, \bibinfo {author}
  {\bibfnamefont {Arun}\ \bibnamefont {Bansil}}, \ and\ \bibinfo {author}
  {\bibfnamefont {Jianwei}\ \bibnamefont {Sun}},\ }\bibfield  {title} {\enquote
  {\bibinfo {title} {Sensitivity of the electronic and magnetic structures of
  cuprate superconductors to density functional approximations},}\ }\href
  {\doibase 10.1038/s41524-022-00711-z} {\bibfield  {journal} {\bibinfo
  {journal} {npj Computational Materials 2022 8:1}\ }\textbf {\bibinfo {volume}
  {8}},\ \bibinfo {pages} {1--11} (\bibinfo {year} {2022})}\BibitemShut
  {NoStop}%
\bibitem [{\citenamefont {Tidey}\ \emph
  {et~al.}(2022{\natexlab{b}})\citenamefont {Tidey}, \citenamefont {Liu},
  \citenamefont {Lai}, \citenamefont {Chuang}, \citenamefont {Chen},
  \citenamefont {Cane}, \citenamefont {Lester}, \citenamefont {Petsch},
  \citenamefont {Herlihy}, \citenamefont {Simonov}, \citenamefont {Hayden},\
  and\ \citenamefont {Senn}}]{Tidey2022a}%
  \BibitemOpen
  \bibfield  {author} {\bibinfo {author} {\bibfnamefont {Jeremiah~P.}\
  \bibnamefont {Tidey}}, \bibinfo {author} {\bibfnamefont {En~Pei}\
  \bibnamefont {Liu}}, \bibinfo {author} {\bibfnamefont {Yen~Chung}\
  \bibnamefont {Lai}}, \bibinfo {author} {\bibfnamefont {Yu~Chun}\ \bibnamefont
  {Chuang}}, \bibinfo {author} {\bibfnamefont {Wei~Tin}\ \bibnamefont {Chen}},
  \bibinfo {author} {\bibfnamefont {Lauren~J.}\ \bibnamefont {Cane}}, \bibinfo
  {author} {\bibfnamefont {Chris}\ \bibnamefont {Lester}}, \bibinfo {author}
  {\bibfnamefont {Alexander~N.D.}\ \bibnamefont {Petsch}}, \bibinfo {author}
  {\bibfnamefont {Anna}\ \bibnamefont {Herlihy}}, \bibinfo {author}
  {\bibfnamefont {Arkadiy}\ \bibnamefont {Simonov}}, \bibinfo {author}
  {\bibfnamefont {Stephen~M.}\ \bibnamefont {Hayden}}, \ and\ \bibinfo {author}
  {\bibfnamefont {Mark}\ \bibnamefont {Senn}},\ }\bibfield  {title} {\enquote
  {\bibinfo {title} {Pronounced interplay between intrinsic phase-coexistence
  and octahedral tilt magnitude in hole-doped lanthanum cuprates},}\ }\href
  {\doibase 10.1038/s41598-022-18574-1} {\bibfield  {journal} {\bibinfo
  {journal} {Scientific Reports}\ }\textbf {\bibinfo {volume} {12}},\ \bibinfo
  {pages} {1--13} (\bibinfo {year} {2022}{\natexlab{b}})}\BibitemShut {NoStop}%
\bibitem [{\citenamefont {Tranquada}\ \emph {et~al.}(2008)\citenamefont
  {Tranquada}, \citenamefont {Gu}, \citenamefont {H\"{u}cker}, \citenamefont
  {Jie}, \citenamefont {Kang}, \citenamefont {Klingeler}, \citenamefont {Li},
  \citenamefont {Tristan}, \citenamefont {Wen}, \citenamefont {Xu},
  \citenamefont {Xu}, \citenamefont {Zhou},\ and\ \citenamefont
  {v.~Zimmermann}}]{Tranquada2008}%
  \BibitemOpen
  \bibfield  {author} {\bibinfo {author} {\bibfnamefont {J.~M.}\ \bibnamefont
  {Tranquada}}, \bibinfo {author} {\bibfnamefont {G.~D.}\ \bibnamefont {Gu}},
  \bibinfo {author} {\bibfnamefont {M.}~\bibnamefont {H\"{u}cker}}, \bibinfo
  {author} {\bibfnamefont {Q.}~\bibnamefont {Jie}}, \bibinfo {author}
  {\bibfnamefont {H.-J.}\ \bibnamefont {Kang}}, \bibinfo {author}
  {\bibfnamefont {R.}~\bibnamefont {Klingeler}}, \bibinfo {author}
  {\bibfnamefont {Q.}~\bibnamefont {Li}}, \bibinfo {author} {\bibfnamefont
  {N.}~\bibnamefont {Tristan}}, \bibinfo {author} {\bibfnamefont {J.~S.}\
  \bibnamefont {Wen}}, \bibinfo {author} {\bibfnamefont {G.~Y.}\ \bibnamefont
  {Xu}}, \bibinfo {author} {\bibfnamefont {Z.~J.}\ \bibnamefont {Xu}}, \bibinfo
  {author} {\bibfnamefont {J.}~\bibnamefont {Zhou}}, \ and\ \bibinfo {author}
  {\bibfnamefont {M.}~\bibnamefont {v.~Zimmermann}},\ }\bibfield  {title}
  {\enquote {\bibinfo {title} {Evidence for unusual superconducting
  correlations coexisting with stripe order in
  {L}a$_{2-x}${B}a$_x${C}u{O}$_4$},}\ }\href {\doibase
  10.1103/physrevb.78.174529} {\bibfield  {journal} {\bibinfo  {journal}
  {Physical Review B}\ }\textbf {\bibinfo {volume} {78}} (\bibinfo {year}
  {2008}),\ 10.1103/physrevb.78.174529}\BibitemShut {NoStop}%
\bibitem [{\citenamefont {Tajima}\ \emph {et~al.}(2001)\citenamefont {Tajima},
  \citenamefont {Noda}, \citenamefont {Eisaki},\ and\ \citenamefont
  {Uchida}}]{Tajima2001}%
  \BibitemOpen
  \bibfield  {author} {\bibinfo {author} {\bibfnamefont {S.}~\bibnamefont
  {Tajima}}, \bibinfo {author} {\bibfnamefont {T.}~\bibnamefont {Noda}},
  \bibinfo {author} {\bibfnamefont {H.}~\bibnamefont {Eisaki}}, \ and\ \bibinfo
  {author} {\bibfnamefont {S.}~\bibnamefont {Uchida}},\ }\bibfield  {title}
  {\enquote {\bibinfo {title} {$c$-{A}xis {O}ptical {R}esponse in the {S}tatic
  {S}tripe {O}rdered {P}hase of the {C}uprates},}\ }\href {\doibase
  10.1103/physrevlett.86.500} {\bibfield  {journal} {\bibinfo  {journal} {Phys.
  Rev. Lett.}\ }\textbf {\bibinfo {volume} {86}},\ \bibinfo {pages} {500--503}
  (\bibinfo {year} {2001})}\BibitemShut {NoStop}%
\bibitem [{\citenamefont {Migliori}\ \emph {et~al.}(1990)\citenamefont
  {Migliori}, \citenamefont {Visscher}, \citenamefont {Brown}, \citenamefont
  {Fisk}, \citenamefont {Cheong}, \citenamefont {Alten}, \citenamefont
  {Ahrens}, \citenamefont {Kubat-Martin}, \citenamefont {Maynard},
  \citenamefont {Huang}, \citenamefont {Kirk}, \citenamefont {Gillis},
  \citenamefont {Kim},\ and\ \citenamefont {Chan}}]{Migliori1990}%
  \BibitemOpen
  \bibfield  {author} {\bibinfo {author} {\bibfnamefont {A.}~\bibnamefont
  {Migliori}}, \bibinfo {author} {\bibfnamefont {William~M.}\ \bibnamefont
  {Visscher}}, \bibinfo {author} {\bibfnamefont {S.~E.}\ \bibnamefont {Brown}},
  \bibinfo {author} {\bibfnamefont {Z.}~\bibnamefont {Fisk}}, \bibinfo {author}
  {\bibfnamefont {S.-W.}\ \bibnamefont {Cheong}}, \bibinfo {author}
  {\bibfnamefont {B.}~\bibnamefont {Alten}}, \bibinfo {author} {\bibfnamefont
  {E.~T.}\ \bibnamefont {Ahrens}}, \bibinfo {author} {\bibfnamefont {K.~A.}\
  \bibnamefont {Kubat-Martin}}, \bibinfo {author} {\bibfnamefont {J.~D.}\
  \bibnamefont {Maynard}}, \bibinfo {author} {\bibfnamefont {Y.}~\bibnamefont
  {Huang}}, \bibinfo {author} {\bibfnamefont {D.~R.}\ \bibnamefont {Kirk}},
  \bibinfo {author} {\bibfnamefont {K.~A.}\ \bibnamefont {Gillis}}, \bibinfo
  {author} {\bibfnamefont {H.~K.}\ \bibnamefont {Kim}}, \ and\ \bibinfo
  {author} {\bibfnamefont {M.~H.~W.}\ \bibnamefont {Chan}},\ }\bibfield
  {title} {\enquote {\bibinfo {title} {Elastic constants and specific-heat
  measurements on single crystals of
  ${\mathrm{la}}_{2}$${\mathrm{cuo}}_{4}$},}\ }\href {\doibase
  10.1103/PhysRevB.41.2098} {\bibfield  {journal} {\bibinfo  {journal} {Phys.
  Rev. B}\ }\textbf {\bibinfo {volume} {41}},\ \bibinfo {pages} {2098--2102}
  (\bibinfo {year} {1990})}\BibitemShut {NoStop}%
\end{thebibliography}%

\section*{Acknowledgements}

M.S.S. acknowledges the Royal Society for a University Research Fellowship
(UF160265). C.K. was supported through a studentship in the Centre for Doctoral
Training on Theory and Simulation of Materials at Imperial College London
funded by the EPSRC (EP/S515085/1 and EP/L015579/1). We acknowledge support
from the Thomas Young Centre under Grant No. TYC-101.  
We are grateful to the UK Materials and Molecular Modelling Hub for
computational resources, which is partially funded by EPSRC (EP/P020194/1
and EP/T022213/1).
We acknowledge the Imperial College London Research Computing Service
(DOI:\url{10.14469/hpc/2232}) for the
computational resources used in carrying out this work.
\newline

\section*{Author contributions}

The work was carried out by C.K. under the supervision of M.S.S., N.C.B. and
A.A.M. The paper was drafted by C.K. with all authors contributing to its
revision.

\section*{Competing interests}

The authors declare no competing interests.

\end{document}